\documentclass{aa}  

\usepackage{natbib}
\bibpunct{(}{)}{;}{a}{}{,} 

\usepackage{array, multirow, graphicx}
\usepackage[dvipsnames]{xcolor}
\usepackage{wasysym}[integrals]
\usepackage{amssymb}
\usepackage{pifont}
\usepackage{siunitx}
\usepackage{booktabs, tabularx}
\usepackage{cellspace, makecell} 
\usepackage{mathtools}
\usepackage{txfonts}

\usepackage[]{hyperref}
\hypersetup{colorlinks,citecolor=blue, linkcolor=blue, urlcolor=blue, breaklinks=true}

\newcounter{magicrownumbers}

\newcommand{\RNum}[1]{\uppercase\expandafter{\romannumeral #1\relax}}
\newcommand{\cii}{[C\RNum{2}]$_{\rm 158\,\mu m}$}
\newcommand{\ci}[1]{[C\RNum{1}]$_{\rm #1\,\mu m}$}
\newcommand{\eq}[1]{\begin{equation} #1 \end{equation}}
\newcommand{\abs}[1]{\left | #1 \right |}

\begin{document} 

   \title{The MUSE Ultra Deep Field (MUDF)}

    \subtitle{VII. Probing high-redshift gas structures in the surroundings of ALMA-identified massive dusty galaxies}

    \titlerunning{MUDF. VII. High-$z$ gaseous structures around massive dusty galaxies}
    
   \author{A. Pensabene\inst{\ref{unimib}}\fnmsep\thanks{\email{antonio.pensabene@unimib.it}}
          \and
          M. Galbiati\inst{\ref{unimib}}\fnmsep\thanks{\email{m.galbiati29@campus.unimib.it}; second author with equal contribution.}
          \and
          M. Fumagalli\inst{\ref{unimib},\ref{oats}}
          \and
          M. Fossati\inst{\ref{unimib},\ref{brera}}
          \and
          I. Smail\inst{\ref{durham}}
          \and
          M. Rafelski\inst{\ref{stsci},\ref{john_hopkins}}
          \and
          M. Revalski\inst{\ref{stsci}}
          \and
          \\F. Arrigoni-Battaia\inst{\ref{mpa}}
          \and
          A. Beckett\inst{\ref{stsci}}
          \and
          S. Cantalupo\inst{\ref{unimib}}
          \and
          R. Dutta\inst{\ref{pune}}
          \and
          E. Lusso\inst{\ref{unifi},\ref{oas}}
          \and
          T. Lazeyras\inst{\ref{unimib}}
          \and
          G. Quadri\inst{\ref{unimib}}
          \and
          D. Tornotti\inst{\ref{unimib}}
          }
          
\institute{Dipartimento di Fisica ``G. Occhialini'', Universit\`a degli Studi di Milano-Bicocca, Piazza della Scienza 3, I-20126 Milano, Italy
\label{unimib}
 \and
    INAF--Osservatorio Astronomico di Trieste, Via G. B. Tiepolo 11, I-34143 Trieste, Italy\label{oats}    
\and
    INAF--Osservatorio Astronomico di Brera, Via Brera 28, I-21021 Milano, Italy\label{brera}
\and
    Centre for Extragalactic Astronomy, Department of Physics, Durham University, South Road, Durham DH1 3LE, UK\label{durham}
\and
    Space Telescope Science Institute, 3700 San Martin Drive, Baltimore, MD 21218, USA\label{stsci}
\and 
    Department of Physics and Astronomy, Johns Hopkins University, Baltimore, MD 21218, USA\label{john_hopkins}
\and
    Max-Planck-Institut f\"{u}r Astrophysik, Karl-Schwarzschild-Str. 1, D-85748 Garching bei M\"{u}nchen, Germany\label{mpa}
\and
    IUCAA, Postbag 4, Ganeshkind, Pune 411007, India\label{pune}
\and
    Dipartimento di Fisica e Astronomia, Universit\`a di Firenze, Via G. Sansone 1, I-50019 Sesto Fiorentino, Firenze, Italy\label{unifi}
\and
    INAF--Osservatorio Astrofisico di Arcetri, Largo Enrico Fermi 5, I-50125 Firenze, Italy\label{oas}
}

   \date{\today}

\abstract{
We present new ALMA continuum and spectral observations of the MUSE Ultra Deep Field (MUDF), a $2\times 2$ arcmin$^2$ region with ultradeep multiwavelength imaging and spectroscopy hosting two bright $z\approx 3.22$ quasars used to study intervening gas structures in absorption.  
Through a blind search for dusty galaxies, we identified a total of seven high-confidence sources, six of which with secure spectroscopic redshifts. We estimate galaxy dust and stellar masses ($M_{\rm dust}\simeq 10^{7.8-8.6}\,M_{\astrosun}$, $M_{\star}\simeq 10^{10.2-10.7}\,M_{\astrosun}$), as well as star formation rates (${\rm SFR}\simeq 10^{1.2-2.0}\,M_{\astrosun}\,{\rm yr^{-1}} $) which show that most of these galaxies are massive and dust-obscured similar to coeval (sub-)millimeter galaxies. All six spectroscopically-confirmed galaxies are within $500~\rm km~s^{-1}$ of metal absorption lines observed in the quasar sightlines, corresponding to $100\%$ association rate. We also find that four of these galaxies belong to groups in which they are among the most massive members. Within the multiple group galaxies associated to the same absorption system, the ALMA sources are not always the closest in projection, but they are often aligned with the gaseous structures in velocity space. This suggest that these massive galaxies occupy the center of the potential well of the gas structures traced in absorption.
However, albeit the low number density of sources identified with ALMA, our study may indicate that absorbers seem to infrequently originate in the inner circumgalactic medium of these galaxies. Instead, they appear to be better tracers of the gas distributed in the large-scale structure that host them.
}


  \maketitle

\section{Introduction}

Absorption lines identified in quasar spectra are excellent tracers of cosmic gas structures across a wide range of redshift and matter overdensities. Sensitive measurements of low-column-density gas against bright quasars have extensively mapped the redshift distribution of neutral and ionized hydrogen as well as that of metals inside the intergalactic medium (IGM) and around galaxies, within the circumgalactic medium (CGM), from the epoch of reionization to present days (see, e.g., \citealt{Oppenheimer+2012, Tumlinson+2017}, and \citealt{Peroux+2020} for a review).   

To date, advances in instrumentation, particularly with integral field spectrographs (IFSs), have enabled a direct view of hydrogen and metal-enriched gas external to the galaxy interstellar medium (ISM) in emission \citep[see, e.g.,][]{Hennawi+2013,Borisova+2016,Cantalupo+2017,Umehata+2019, Bacon+2021, Leclercq+2022, Dutta+2024b, Tornotti+2024}. However, these studies are limited to the IGM and CGM densest portions, typically at close separations from galaxies. Therefore, absorption spectroscopy in quasar fields remains a crucial way to unravel the connection between the galaxy physical properties (e.g., stellar mass, star-formation rate, morphology) and those processes that regulate galaxy formation and evolution (e.g., gas accretion and outflows) within their larger-scale environment \citep[see, e.g.,][]{Weiner+2009, Chen+2010a, Chen+2010b, Steidel+2010, Zabl+2019, Dutta+2020, Galbiati+2023, Schroetter+2024}. 

At observed optical wavelengths, recent works in a wide range of redshifts, from $z\approx 0.5$ to $z\approx 4.5$, revealed a clear clustering of metal absorption lines, such as \ion{Mg}{II} and \ion{C}{IV} absorbers \citep[see, e.g.,][]{Dutta+2021, Banerjee+2023, Galbiati+2023, Galbiati+2024}, or strong hydrogen absorbers, i.e., the Lyman limit systems (LLSs) and damped Ly$\alpha$ absorbers \citep[DLAs;][]{Fumagalli+2016, Lofthouse+2023} around star-forming galaxies. These dense and complete spectroscopic surveys highlighted the importance of the galaxy environment in shaping the gas distribution at all redshifts, with a clear excess of cool/warm gas seen in absorption near group galaxies compared to isolated ones. With the clustering of multiple galaxies near absorption systems, a more complex picture emerges, where at least a fraction of the absorbers are probing the gas-rich cosmic structures within which galaxies reside. 

At longer wavelengths, surveys in the (sub-)millimeter, especially with the Atacama Large Millimeter Array (ALMA), provided a complementary view of the dusty and gas-rich galaxies associated with absorption line systems in quasar fields. Following the first detection of \cii{}, line and dust-continuum emission from galaxies associated with $z\approx 4$ DLAs \citep{Neeleman+2017}, systematic searches of molecular lines at $z\approx 0.5-2.5$ in the proximity of high-column density \ion{H}{I} systems show that the highest-metallicity DLAs are associated with the most massive galaxies \citep{Moller+2018, Kanekar+2018, Kanekar+2020, Neeleman+2018, Neeleman+2019, Kaur+2022, Kaur+2022b}. As in the case of optical searches, (sub-)millimeter observations uncover a wide range of environments, from more isolated galaxies to groups. 

Based on the above results, multiple tracers are needed to complement our view of the gas distribution around cosmic structures populated by galaxies. Indeed, efforts to combine galaxy observations in quasar fields across wavelengths, especially with the Multi Unit Spectroscopic Explorer \citep[MUSE;][]{Bacon+2010} at the Very Large Telescope (VLT) and ALMA, are multiplying \citep[e.g.,][]{Klitsch+2018,Peroux+2019, Szakacs+2021}. These combined observations reveal examples of groups composed of optically selected galaxies and dusty or molecular-rich galaxies near strong hydrogen absorbers \citep{Fynbo+2018, Klitsch+2018, Klitsch+2019}, particularly LLSs and DLAs. Conversely, multi-wavelength searches for galaxies associated with metal lines have been explored less \citep[see, e.g.,][]{Kashino+2023}. However, results similar to those of LLSs and DLAs are expected. For instance, the search of Ly$\alpha$ emitting galaxies near \ion{Mg}{II} absorbers with MUSE reveals an unexpected lack of detections near high-equivalent width absorbers at $z\approx 3-4$, perhaps hinting at the presence of an unseen dusty galaxy population \citep{Galbiati+2024}.

Following this line of inquiry, in this paper, we present new ALMA observations of the MUSE Ultra Deep Field \citep[MUDF;][]{Lusso+2019, Fossati+2019, Revalski+2023}, a unique set of multiwavelength imaging and wide-field spectroscopic observations (from UV to near IR) at exceptional depths within a $2\times 2$~arcmin$^2$ field hosting two bright $z\approx 3.22$ quasars (Q2139-4433 and Q2139-4434, hereafter QSO-NW and QSO-SE, respectively), which we use to map in absorption gaseous cosmic structures at $z\lesssim 3.22$. In this paper, we extend the MUDF dataset to longer wavelengths by mapping the 1.2-mm dust continuum with ALMA band 6 and spectral scans in band 3 to target the rotational transitions of carbon monoxide CO. With these data, we study the association of dusty and gas-rich galaxies with metal absorbers detected in the MUDF volume without preselection on the absorber properties. Comparisons with detections at optical and near-infrared (NIR) wavelengths, reaching the dwarf galaxy regime, are also presented to build a complete view of the galaxy environment surrounding absorbers.

This paper is structured as follows: in Sect.~\ref{sect:data-red} we present the acquired data and illustrate the data processing. In Sect.~\ref{sect:ancillary_datasets}, we describe the ancillary datasets used in this work. We devote Sect.~\ref{sect:source_extraction} to analyzing the data, extracting the sources, and measuring the flux continuum and line luminosities. In Sect.~\ref{sect:results_discussion}, we discuss the properties of our galaxy sample in connection with those of the surrounding gas detected in absorption along the quasar sightlines. Finally, in Sect.~\ref{sect:summary_conclusions} we summarize the results and draw our conclusions. In this work, we assume a standard $\Lambda$ cold dark matter cosmology with $H_0 = 67.7\,{\rm km\,s^{-1}\,Mpc^{-1}}$, $\Omega_{\rm m} = 0.310$, and $\Omega_\Lambda = 1 - \Omega_{\rm m}$ from \citet{PlanckColl+2020}. We use AB magnitude and adopt $3\sigma$ limits when not stated otherwise.

   \begin{figure*}[!ht]
    	\centering
   	\resizebox{\hsize}{!}{
		\includegraphics{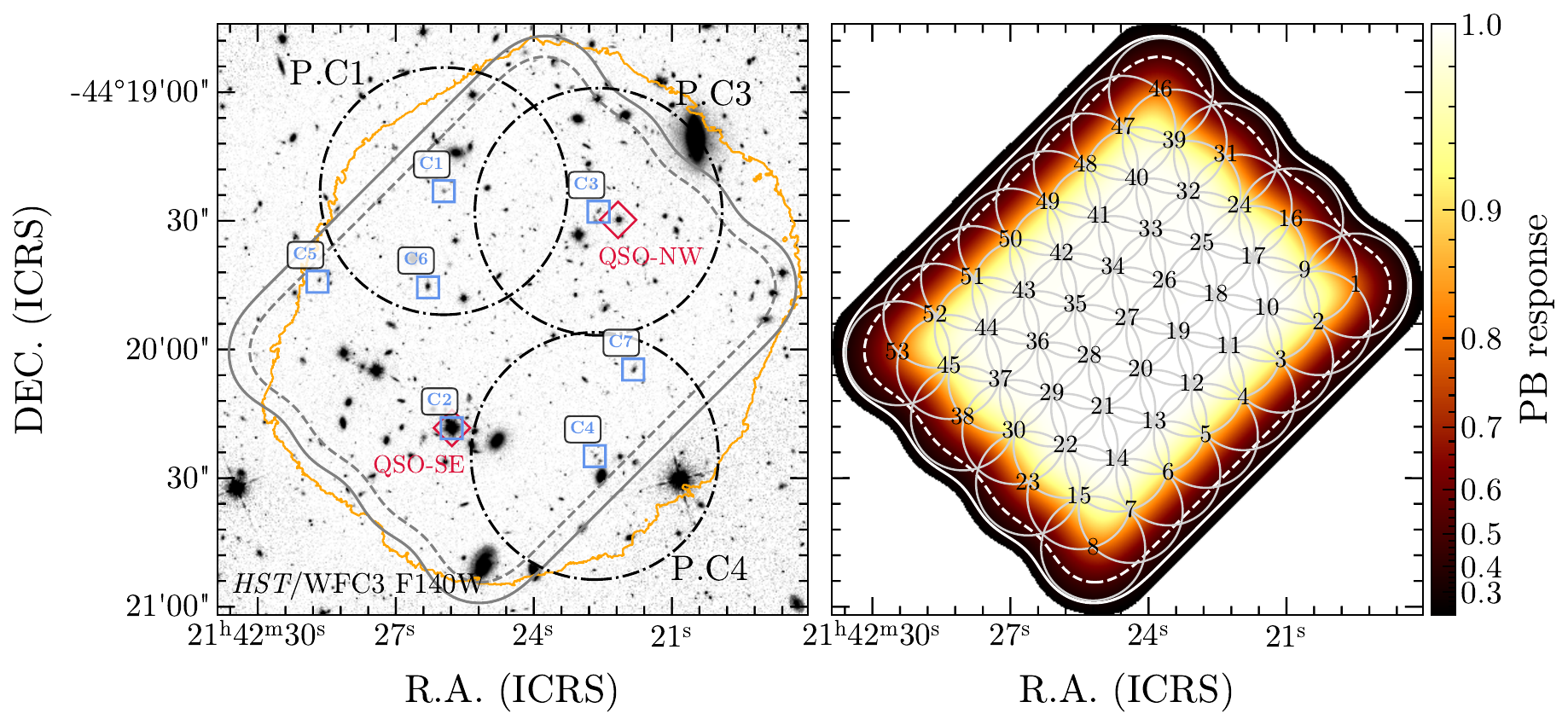}}
       \caption{Footprints of ALMA observations toward the MUDF. {\it Left panel:} The background shows the HST/WFC3 F140W image, and the gray lines show the PB response of the ALMA band 6 mosaic at $30\%$ and $50\%$ (solid and dashed lines, respectively). The orange line indicates the region observed by VLT/MUSE with $>3\,{\rm hours}$ of exposure time. Red diamonds indicate the MUDF quasar pair.
       The blue squares with labels indicate the high-fidelity ALMA-selected sources extracted from the $1.2\,{\rm mm}$ continuum image. The dot-dashed black circles show the ALMA pointings of the band 3 spectral scans toward three individual sources (P.C1, P.C3, and P.C4). The diameter of the circles indicates the HPBW at $101\,{\rm GHz}$. {\it Right panel:} PB response of the ALMA band 6 mosaic. The circles show the disposition of the 53 pointings with a diameter equal to the HPBW of the ALMA $12\,{\rm m}$ antennas at the reference frequency of the setup.}
       \label{fig:pb_pointings}
   \end{figure*}

    \begin{figure}[!ht]
    	\centering
   	\resizebox{\hsize}{!}{
		\includegraphics{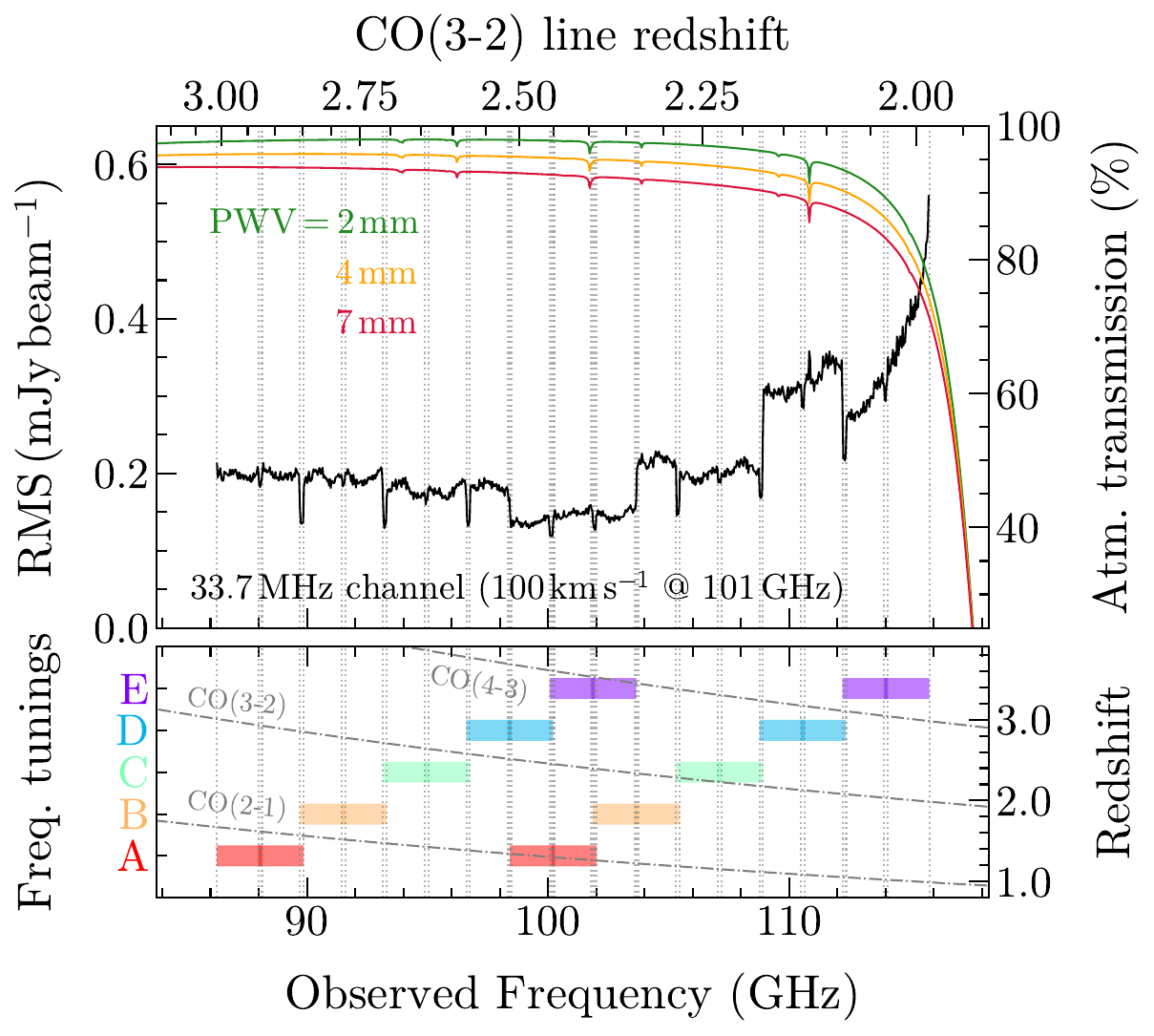}}
       \caption{Sensitivity (RMS) of the ALMA band 3 spectral scans across the covered frequency range. {\it Top panel:} The colored curves and the right axis report the atmospheric transmission from the ALMA site at different levels of PWV obtained using {\tt plotAtmosphere} task in the CASA {\tt analysisUtils} package \citep{Hunter+2023b}. The top axis reports the expected redshift of the CO(3--2) line. {\it Bottom panel:} Disposition of SPWs in the four frequency tunings. Vertical dotted lines indicate the edges of the SPWs. The maximum sensitivity is achieved within $z\approx2.35-2.5$ where two SPWs in different frequency tunings overlap. The right axis reports the redshift of the various CO transitions (dash-dotted lines) covered by the spectral scan within $z\simeq1-4$. }
       \label{fig:sens_scans}
   \end{figure}

\section{ALMA observations and data reduction}
\label{sect:data-red}
\subsection{Description of observations}
In this work, we use ALMA band 6 mosaic observations from the program 2021.1.00285.S (Cycle 8, PI: M. Fumagalli), along with three ALMA band 3 spectral scans toward individual sources within the sky region covered by the band 6 mosaic (program ID: 2023.1.00461.S, Cycle 10, PI: A. Pensabene). The latter observations were designed to accurately measure the redshift of sources lacking robust spectroscopic constraints from MUSE (see Sect.~\ref{ssect:hst_counterparts}). In what follows, we describe the acquisition and processing of the data.

The ALMA band 6 mosaic has been designed to cover the central part of the MUDF, where the MUSE observations achieve the maximum depth \citep[see,][]{Tornotti+2024}. The mosaic consists of 53 Nyquist-sampled pointings, each with a half-power beam width (HPBW) of $\approx24\rlap{.}{\arcsec}25$ at the reference frequency of $240.10\,{\rm GHz}$, resulting in a covered rectangular sky area of $\approx1\rlap{.}{\arcmin}9\times1\rlap{.}{\arcmin}6$ (Fig.~\ref{fig:pb_pointings}). Observations were carried out in frequency division mode (FDM). The spectral tuning is composed of four $1.875\,{\rm GHz}$-wide spectral windows (SPWs) centered at frequencies $254.498\,{\rm GHz}$, and $257.998\,{\rm GHz}$ in the upper sideband (USB), and $240.102\,{\rm GHz}$ and $243.102\,{\rm GHz}$ in the lower sideband (LSB), respectively. The total effective bandwidth of the spectral setup is $7.5\,{\rm GHz}$ centered at wavelength $1.2\,{\rm mm}$.  Observations were subdivided into five execution blocks (EBs) carried out during the period from May 28 to June 3, 2022 employing 41 to 44 12-m antennas with baselines ranging between $15.1- 783.5\,{\rm m}$ yielding a naturally-weighted synthesized beam FWHM of $0\rlap{.}{\arcsec}55\times0\rlap{.}{\arcsec}51$, and were conducted with a mean precipitable water vapor (PWV) of $0.3 - 1\,{\rm mm}$. The total on-source integration time was $\simeq 5.86\,{\rm h}$ achieving a sensitivity of $26\,{\rm \mu Jy\,beam^{-1}}$ over the total bandwidth. The native spectral resolution is $\simeq 7.8\,{\rm MHz}$, corresponding to $\simeq 9.5\,{\rm km\,s^{-1}}$ at the reference frequency of $240.1\,{\rm GHz}$. The J2126-4605 and J2258-2758 were used as phase and flux calibrators during observations.

The ALMA band 3 observations consist of three single-pointing spectral scans centered on three sources revealed in the band 6 mosaic (see, Fig.~\ref{fig:pb_pointings}) to target their CO(3--2) line (rest-frame frequency $\nu_{\rm rest}=461.041\,{\rm GHz}$) in the redshift range $z\simeq 2-3$. The frequency setups were tuned based on the photometric redshift probability distributions (pointing to the redshift range $1.8\lesssim z \lesssim 2.8$) obtained from multiwavelength photometry available for galaxies in the MUDF \citep[see,][]{Revalski+2023} using the {\sc EAZY} code \citep{Brammer+2008}.  Observations were carried out in FDM employing five frequency tunings with two adjacent $1.875\,{\rm GHz}$-wide SPWs per sideband designed to cover the spectral range $86.25 - 115.80\,{\rm GHz}$ contiguously. The total effective bandwidth is $29.55\,{\rm GHz}$. Each frequency tuning was observed in two EBs, and executed between January 18th and March 4th 2024, employing 40 to 45 12-m antennas with baselines ranging in $15.1-483.9\,{\rm m}$ and a mean ${\rm PWV}=2.7-7.1\,{\rm mm}$. The $uv$ coverage of these observations yields to a naturally-weighted synthesized beam size of $2\rlap{.}{\arcsec}5\times2\rlap{.}{\arcsec}3$ at the reference frequency of $101\,{\rm GHz}$. The total on-source observing time was $8.1\,{\rm h}$. Data were acquired with a native spectral resolution of $1.95\,{\rm MHz}$, corresponding to $\simeq5.8\,{\rm km\,s^{-1}}$ at $101\,{\rm GHz}$. The sensitivity achieved in the final dataset varies along with frequency due to the variation in atmospheric absorption and weather conditions across the EBs (Fig.~\ref{fig:sens_scans}) and has a median value of $0.20\,{\rm mJy\,beam^{-1}}$ over $33.7\,{\rm MHz}$, or $\simeq 100\,{\rm km\,s^{-1}}$ at $101\,{\rm GHz}$. Fig.~\ref{fig:pb_pointings} shows the footprints of the ALMA observations presented in this work, the pointing design, and the response of the primary beam (PB) of the band 6 mosaic. In Fig.~\ref{fig:sens_scans}, we show the spectral coverage of the acquired band 3 spectral scans and the sensitivity variation along the frequency range. 

\subsection{Data calibration and imaging}
\label{ssect:data_calibration}
We calibrated the data using the Common Astronomy Software Application \citep[CASA;][]{McMullin+2007, Hunter+2023} by running the calibration pipeline \texttt{scriptForPI} provided with the raw measurement set using the CASA pipeline versions 6.2.1 and 6.5.4 for band 6 and 3 data, respectively. No self-calibration was performed on the datasets. We imaged band 6 visibilities by using the CASA task {\tt tclean}. We first obtained a ``dirty'' continuum image by performing the Fourier transform of the visibilities ({\tt niter=0}) using the multi-frequency synthesis mode ({\tt specmode = ``mfs''}) and adopting a natural weighting scheme to maximize the sensitivity per beam. In this procedure, we set ``mosaic'' as the gridding convolution function and put the phase center at the coordinates ICRS 21:42:24.4658 $-$44:19:53.007. We used a pixel scale of $0\rlap{.}{\arcsec}08$ to achieve the Nyquist sampling of the longest baselines. Similarly, we obtained ``dirty'' band 6 data cubes of the LSB and USB with $25\,{\rm km\,s^{-1}}$ channel width. {Additionally, in order to improve the surface brightness sensitivity for faint extended emission, we obtained continuum image by applying a Gaussian taper in the $uv$ plane with ${\rm FWHM}_{uv}=90\,{\rm k}\lambda$ and setting the pixel size to $0\rlap{.}{\arcsec}4$, yielding a synthesized beam FWHM of $2\rlap{.}{\arcsec}19\times2\rlap{.}{\arcsec}39$.} The ``dirty'' data preserve the intrinsic properties of the noise, and we used that to perform a blind search of continuum and line emitters in the field (see Sect.~\ref{sect:source_extraction}). We finally obtained ``cleaned'' continuum images and cubes by applying the cleaning algorithm down to $2\sigma$, placing circular masks of $2\rlap{.}{\arcsec}0$ diameter around sources extracted with $S/N>5$.

We imaged the ALMA band 3 pointings by obtaining ``dirty'' cubes of the combined dataset, including all the SPWs using {\tt tclean} with {\tt niter=0} and the ``natural'' weighting scheme of the visibilities. We set the pixel scale at $0\rlap{.}{\arcsec}3$ and the channel width at $100\,{\rm km\,s^{-1}}$ at the reference frequency of $101\,{\rm GHz}$, which is expected to maximize the $S/N$ per channel while resolving the line profile. In such datasets, sources are located at the phase center and appear faint and spatially unresolved and no bright continuum emitters are found in the cubes. For these reasons, the cleaning of cubes is not required. 

\section{Ancillary multi-wavelength observations}
\label{sect:ancillary_datasets}
In this study, we use the catalog of MUDF galaxies from \citet{Revalski+2023}, which provides deep photometry and high-confidence redshifts constrained with HST and MUSE spectroscopy, to associate optical counterparts to the sources identified by ALMA. This field benefits from extremely deep MUSE observations (ESO PID 1100.A-0528, PI: M. Fumagalli) with an integration time of $\approx$142 hours in wide field mode and with extended wavelength coverage ($4600-9350\,\text{\AA}$, and spectral resolution $R\approx2000-4000$). We refer to \citet{Fossati+2019} for detailed descriptions of the survey design and the data reduction. 

The MUDF survey is specifically designed to study the connection between gas and galaxies across cosmic time and thus also includes high-resolution ($R\approx40,000$) spectra of the two quasars (QSO-SE and QSO-NW with $r$-band magnitude 17.9 and 20.5, respectively) to probe the IGM and CGM in absorption \citep[see,][]{Beckett+2024}. The quasar spectra were obtained using the Ultraviolet and Visual Echelle Spectrograph (UVES, \citealp{Dekker+2000}) at the VLT (PIDs: 65.O-0299, 68.A-0216, 69.A-0204, and 102.A-0194, PI: V. D'Odorico) and cover the wavelength range $4100-9000\,\text{\AA}$ with a $S/N$ per pixel of $\approx25$ and $\approx10$ for the bright and faint quasar, respectively.

In HST Cycle 26, we also collected 90 orbits of WFC3/F140W imaging and WFC3/IR G141 grism spectroscopy (PID 15637, PI: M. Rafelski \& M. Fumagalli), providing a deep and extensive sampling of the galaxy spectral energy distributions (SEDs) from UV to NIR wavelengths \citep[see,][]{Revalski+2024}. Custom calibration of HST observations is described by \citet{Revalski+2023} together with a complete list of additional optical and near-UV (NUV) photometry\footnote{The final products can be retrieved as High Level Science Products (HLSPs) at {\tt doi:}\href{https://doi.org/10.17909/81fp-2g44}{10.17909/81fp-2g44}, together with the default data files at {\tt doi:}\href{https://doi.org/10.17909/q67p-ym16}{10.17909/q67p-ym16}
from the Mikulski Archive for Space Telescopes (MAST). The dedicated page to the MUDF program is available at \url{https://archive.stsci.edu/hlsp/mudf}.}. 

\subsection{Catalog of HST-selected galaxies}\label{ssect:hst_cat}
Galaxies detected in the deep WFC3/F140W image are included in a catalog that is $50\%$ complete at $27.6\rm\,mag$ and covers the entire area observed with MUSE, as described by \citet{Revalski+2023}.  To all the detected galaxies, a spectroscopic redshift is assigned together with a quality flag (see table 5 in \citealp{Revalski+2023}), depending on the presence of high $S/N$ lines in their spectra. The spectral fitting process used to identify and fit emission lines across a wide range of wavelengths, from UV to NIR, is detailed in \citet{Revalski+2024}. A high-quality redshift (redshift quality flag $\geq3$) was assigned to $25\%$ of the sources with spectral coverage. The physical properties of these galaxies, such as stellar mass ($M_{\star}$) and star formation rate (SFR), are obtained by fitting the multiwavelength photometry and the MUSE spectra simultaneously, as described in \citet[][see, also, \citealt{Revalski+2024}]{Fossati+2018} , to which we refer for full details about the method. Finally, we distinguished between isolated galaxies and those residing in groups. To this end, we applied a friends-of-friends algorithm to identify galaxies belonging to a group as all the sources that are not isolated within $\pm 500\,{\rm km\,s^{-1}}$ along the line of sight \citep[see, e.g.,][for a similar approach]{Fossati+2019}. Overall, we found that $\approx86\%$ of the galaxies reside in 23 different groups across the redshift range $z=1-3$, with five groups with more than five galaxies, and two with more than 15 members.

\subsection{Catalog of quasar absorbers}
\label{ssect:qso_absorbers}
The detailed procedure used to identify the absorption line systems in the spectra of the two quasars, as well as the complete list of the transitions and their properties, is illustrated in table 1 of \citet{Beckett+2024}, and is only briefly summarized below. The spectra are normalized by fitting a cubic spline to the continuum redward to the Ly$\alpha$ emission of the quasars by using the ESPRESSO data analysis software \citep{Cupani+2016}. The spectra are then visually inspected to identify absorption line systems such as \ion{Mg}{I}, \ion{Mg}{II}, \ion{Fe}{II}, \ion{Al}{II}, \ion{Al}{III}, \ion{C}{IV} and \ion{Si}{IV}. All these lines are then fitted using the Bayesian code Monte Carlo Absorption Line Fitter (MC-ALF, \citealp{Longobardi+2023}) and decomposed into the minimum number of Voigt profiles required to reproduce the observed flux. The Doppler parameter, column density, and redshift of each Voigt component are the free parameters of the fit. Additional filler components account for blends with absorption features arising from different ions at different redshifts. We finally assign the redshift of the component with the largest column density to each absorber. The resulting sample includes 22 absorption systems in the QSO-SE sightline, and nine systems in the QSO-NW sightline. Only one absorption system is detected in both the sightlines at $z\approx0.88$ (see, table 1 in \citealt{Beckett+2024}). 
Multiple ions are often detected at the redshift of each system.

\section{Extraction and characterization of ALMA sources}
\label{sect:source_extraction}
We performed a source search in the ``dirty'' ALMA band 6 continuum mosaic image using the Python-based code {\sc LineSeeker}\footnote{The code is publicly available at \url{https:// github.com/jigonzal/LineSeeker}. Other similar source-finding codes are available in the literature, such as {\sc FindClumps} \citep{Walter+2016} or MF3D \citep{Pavesi+2018}, which differ mainly in details about the adopted spectral filter function. However, comparisons between codes have yielded very similar results \citep[see][]{Gonzalez-Lopez+2019}.}. The complete details of the codes and performances are described in \citet{Gonzalez-Lopez+2017, Gonzalez-Lopez+2019}. Here, we summarize the basic characteristics and operation of the algorithm. 

{\sc LineSeeker} has been designed to perform a blind search of line emitters in large data cubes from (sub-)millimeter surveys such as the ALMA Spectroscopic Survey in the {\it Hubble} Ultra Deep Field \citep[ASPECS; see, e.g.,][]{Decarli+2019c, Gonzalez-Lopez+2019, Gonzalez-Lopez+2020}, but can be equivalently employed to search sources in 2D continuum images. The code searches for the line emitter in the cube by adopting a matched filter approach. {\sc LineSeeker} convolves the data cube along the spectral axis with a Gaussian kernel with a range of input widths set based on a typical linewidth. For each convolution iteration, the code finds all the voxels above a given $S/N$ threshold. The noise is estimated in the convolved cube through $5\sigma$-clipped statistics, and the signal is measured from the peak flux density per beam. The sources are extracted by grouping connected voxels using the Density-Based Spatial Clustering of Application with Noise algorithm \citep[DBSCAN;][]{Ester+1996}\footnote{This algorithm is included within the Python package Scikit-learn \citep{Pedregosa+2011}.}. Then, the final list of sources is drawn by cross-matching the source catalogs obtained from different convolution kernels and selecting the maximum $S/N$. For each identified candidate, {\sc LineSeeker} estimates the probability of false positive detection by performing the source search on the negative cube. Since any negative peak in a (sub-)millimeter extragalactic survey is expected to be result of noise, the ratio between positive and negative detection distributions is used to estimate the fidelity (or reliability) of source candidates as a function of $S/N$ as
\eq{F(S/N)=1-\frac{N_{\rm neg}(S/N)}{N_{\rm pos}(S/N)},}
where $N_{\rm neg}$ and $N_{\rm pos}$ are the number of negative and positive detections at a given $S/N$. The cumulative distribution of negative detections is then modeled assuming Gaussian-distributed noise by using a function of the form $N[1-{\rm erf}(S/N/\sqrt{2}\,\sigma)$, where erf is the error function, and $N$, $\sigma$ are free parameters. The best-fit model is finally used to estimate $F(S/N)$ for all positive source candidates. The search of continuum sources in 2D images can be performed similarly with {\sc LineSeeker} by skipping the convolution steps. 

\begin{table*}[!t]
\def\arraystretch{1.15}
\caption{ALMA 1.2-mm continuum-selected sources.}  
\label{tbl:1.2mm_cand}      
\centering 
\resizebox{\hsize}{!}{
\begin{tabular}{lccccccccc}\toprule\toprule
ID$_{\rm 1.2mm}\,^{(1)}$ & R.A.$\,^{(2)}$ & DEC.$\,^{(3)}$ & $S_{\rm 1.2\,mm}^{\rm peak}$$\,^{(4)}$ & ${\rm S/N}$$\,^{(5)}$ & fidelity$\,^{(6)}$ &$z_{\rm MUSE}$$\,^{(7)}$ &$z_{\rm ALMA}$$\,^{(8)}$ & ALMA line$\,^{(9)}$ & $m_{\rm F140W}$$\,^{(10)}$\\
	 	& (ICRS) & (ICRS) & $({\rm mJy\,beam^{-1}})$ & & & & & & (mag) \\
\cmidrule(lr){1-10}

C1& 21:42:25.957 & -44:19:23.09 & $0.85\pm0.07$ & $24.3$ & $1.00$ & $-$ & $2.2529$ &CO(3--2) & $23.89$\\
C2 (QSO-SE)& 21:42:25.786 & -44:20:18.37 & $0.75\pm0.06$ & $20.3$ & $1.00$ & $3.2236$ & $3.2267$ &CO(4--3) &$17.34$\\
C3 & 21:42:22.595 & -44:19:27.89 & $0.32\pm0.06$ & $8.4$ & $1.00$ & $-$ & $2.2954$ &CO(3--2) & $23.51$\\
C4& 21:42:22.676 & -44:20:24.85 & $0.47\pm0.08$ & $8.4$ & $1.00$ & $-$ & $2.3676$ &CO(3--2) & $23.81$\\
C5  & 21:42:28.701 & -44:19:44.20 & $0.39\pm0.09$ & $5.6$ & $1.00$ & $2.4723^{(\dagger)}$ & $-$ & $-$ & $23.35$\\
C6& 21:42:26.300 & -44:19:45.49 & $0.27\pm0.06$ & $5.1$ & $0.96\pm0.02$ & $1.0541$ & $1.0544$ &\ci{609} &  $22.21$\\
C7& 21:42:21.841 & -44:20:04.69 & $0.13\pm0.03$ & $5.0$ & $0.94\pm0.02$ & $1.7561$ & $-$ & $-$ &  $22.86$\\

\bottomrule
\end{tabular}
}
\tablefoot{High-fidelity ($F\ge90\%$) candidate list extracted via blind search using {\sc LineSeeker}. $^{(1)}$Identifier of ALMA 1.2-mm continuum-selected candidates. $^{(2)}$Right Ascension (ICRS). $^{(3)}$Declination (ICRS). $^{(4)}$Flux density extracted from the source peak pixel in the $1.2\,{\rm mm}$ tapered continuum image. $^{(5)}$Signal-to-noise ratio. $^{(6)}$Fidelity and its uncertainties estimated by {\sc LineSeeker} using negative detections. All sources have a NIR counterpart, as revealed in {\it HST} / WFC3 F140W image, therefore, the actual fidelity should be considered $F=1$. $^{(7)}$High-confidence redshift measurement from MUSE spectroscopic data. $^{(\dagger)}$Tentative redshift estimate measured from low-quality MUSE spectrum. $^{(8)}$Redshift measurement derived from emission line detected with ALMA (see, also, Table~\ref{tbl:co32_line}). $^{(9)}$Emission line detected in either ALMA band 6 data cube or band 3 spectral scans. $^{(10)}$Apparent AB magnitude in {\it HST}/WFC3 F140W filter. All reported uncertainties correspond to $1\sigma$.}
\end{table*}

\subsection{Continuum-selected sources in the 1.2-mm image}
\label{ssect:cont_selected_sources}
We selected source candidates in the ALMA band 6 mosaic by running {\sc LineSeeker} on the untapered ``dirty'' continuum image, excluding the region below $50\%$ of the maximum PB response, where the low telescope sensitivity increases the probability of selecting spurious sources. The ``dirty'' data are preferred over the ``cleaned'' ones, since the former preserves the intrinsic properties of the noise. Furthermore, we input the ``dirty'' image into {\sc LineSeeker} without correcting for the PB response, thus preserving the spatial homogeneity of the noise across the field-of-view (FoV). With {\sc LineSeeker} we extracted all the detections above $S/N>3$ and selected the source candidates with $F\ge 90\%$, corresponding to $S/N\ge 5.0$. We recovered a total of seven 1.2-mm continuum-selected candidates (labeled C1 to C7), including the host galaxy of the QSO-SE (C2). We report the location of these sources in Fig.\ref{fig:pb_pointings} and their characteristics in Table~\ref{tbl:1.2mm_cand}. We performed similar search on the tapered band 6 continuum image, which is more sensitive to faint extended emission, and found consistent result.

To further investigate the reliability of faint candidates, we complemented the search for continuum sources by cross-matching the catalog of galaxies detected in the HST/WFC3 F140W image (see Sect.~\ref{ssect:hst_cat}) with the sources selected by {\sc LineSeeker} with fidelity values within $20\%\le F <90\%$. To this purpose, we search HST counterparts around the sky position of ALMA continuum sources within a separation limit of $0\rlap{.}{\arcsec}5$, comparable to the beam size of the 1.2-mm continuum untapered image, and found no associations. For the goal of this study, we therefore include in our fiducial sample of ALMA continuum sources only the high-fidelity candidates selected as previously described. For all of these sources, we found associations in the HST catalog as described in Sect~\ref{ssect:hst_counterparts}.

We evaluated the completeness of our ALMA band 6 survey by injecting artificial point-like sources at random locations in the untapered dirty 1.2-mm continuum image. Then, we computed the rate of recovered sources by running {\sc LineSeeker} and searching for candidates with $F\ge90\%$ encompassed within $0\rlap{.}{\arcsec}275$ from the injected position, corresponding to half of the beam major axis. We repeated the simulation by injecting $20$ sources simultaneously into the image and repeating this procedure for different values of flux density in the range $[2 , 10]\times{\rm RMS}$. We finally iterated the entire simulation $30$ times to achieve good statistics. As a result, we find that our source search is complete to $\approx90\%$ at $S/N\approx7.5$. During the simulation, we also evaluated the effect of flux boosting \citep[see, e.g.,][]{Hogg+1998, Scott+2002, Coppin+2006} on the recovered sources, and we determined that it is negligible at $S/N \gtrsim 5$. Thus, we do not correct the measured fluxes of our source sample for this effect.

\subsection{Line emitters in band 6 data cubes}
\label{ssect:line_selected_sources}
The band 6 data cubes cover multiple molecular and atomic lines which enter the LSB and USB at different redshifts. In particular, for $z<3.5$ the carbon monoxide CO transitions are covered by our frequency setup from the CO(3--2) at $z\simeq 0.45$ in the USB, to the CO(9--8) at $z\simeq 3.34$\footnote{Higher-$J$ CO transitions are covered for $z>3.5$ by our ALMA band 6 spectral tunings.}, while the fine-structure lines of the atomic carbon  \ci{609, 369} are observable from $z\approx0.9$ up to $2.38$.

To complete the source search in the field, we employed {\sc LineSeeker} to detect line emitters in the cubes blindly. Similarly to the search for continuum emitters (see Sect.~\ref{ssect:cont_selected_sources}), we restricted the source extractions to the survey area with PB response $\ge 50\%$, and we run the line-search algorithm on both LSB and USB cubes using a Gaussian kernel with width ranging from 1 to 18 channels, which is adequate to match the typical linewidths of ${\rm FWHM}\sim 50-1000\,{\rm km\,s^{-1}}$ observed in high-redshift galaxies \citep[see, e.g.,][]{Birkin+2021}. We extracted all line-emitting candidates with $S/N\ge 3.0$ and then selected those with $F\ge90\%$. As a result, we identified one line emission in the spectrum of the C6 source, which we already identified through its continuum at 1.2 mm (see Table~\ref{tbl:1.2mm_cand}). This source has a robust spectroscopic redshift of $z=1.05413$ measured with the MUSE data (see Sect.~\ref{ssect:hst_counterparts}) that allowed us to identify the line as \ci{609}. We measured the line flux by fitting the spectrum extracted from the source peak pixel (see, Sect.~\ref{ssect:source_fluxes}), and obtained $F_{\rm [CI]609\,\mu m}=0.36\pm0.10\,{\rm Jy\,beam^{-1}\,km\,s^{-1}}$. The source redshift we estimated with this line is consistent with the one measured from the MUSE spectrum. The visual inspection of ALMA band 6 spectra at the location of all the 1.2-mm continuum-selected sources did not reveal additional emission lines, consistently with the spectroscopic redshift measurements of the sources.

Finally, we complemented our line search by cross-matching the HST catalog of galaxies with the source catalog of line-emitting candidates delivered by {\sc LineSeeker} following the same approach described in Sect.~\ref{ssect:cont_selected_sources}. This procedure did not reveal additional line emitters.

\subsection{Source detection in band 3 spectral scans}
\label{ssect:search_spectral_scans}
We analyzed ALMA band 3 spectral scans targeting sources C1, C3, and C4. We inspected the spectra at the center of the pointing around the nominal locations of the continuum detections revealed at 1.2 mm in the ALMA mosaic (see Table~\ref{tbl:1.2mm_cand}). The CO(3--2) line, which is identified based on the posterior distributions of the source photometric redshift\footnote{We derived photometric redshifts of C1, C3 and C4 from multiband photometry as detailed in Sect.~\ref{sect:data-red}, obtaining $z_{\rm phot} = 2.21\pm0.19$, $2.51\pm0.07$, $2.44\pm0.17$, respectively, where nominal values and uncertainties are computed from the expected value and variance of the photo-$z$ posteriors.}, is detected in all three sources, which appear spatially unresolved given the low angular resolution of these ALMA observations (see Sect.~\ref{sect:data-red}). 

We ran {\sc LineSeeker} on band 3 "dirty" cubes and continuum images similarly to what described in Sects.~\ref{ssect:cont_selected_sources} and ~\ref{ssect:line_selected_sources} for band 6 data.  The search for line emitters within the cubes reveal the CO(4--3) line of the QSO-SE (C2) which is redshifted in the ALMA band 3 at $z\approx3.22$. The analysis of this detection is beyond the scope of this work. Additionally, we performed visual inspection of the band 3 spectra in the location of secure continuum sources selected in the ALMA band 6 mosaic, and found no evident emission lines. Finally, in the 3-mm continuum images we did not find neither high-fidelity source candidates, nor HST counterparts associated to the low-fidelity ones.

\subsection{HST counterparts of ALMA continuum sources}
\label{ssect:hst_counterparts}
\begin{figure}[!t]
    	\centering
   	\resizebox{0.95\hsize}{!}{
	\includegraphics{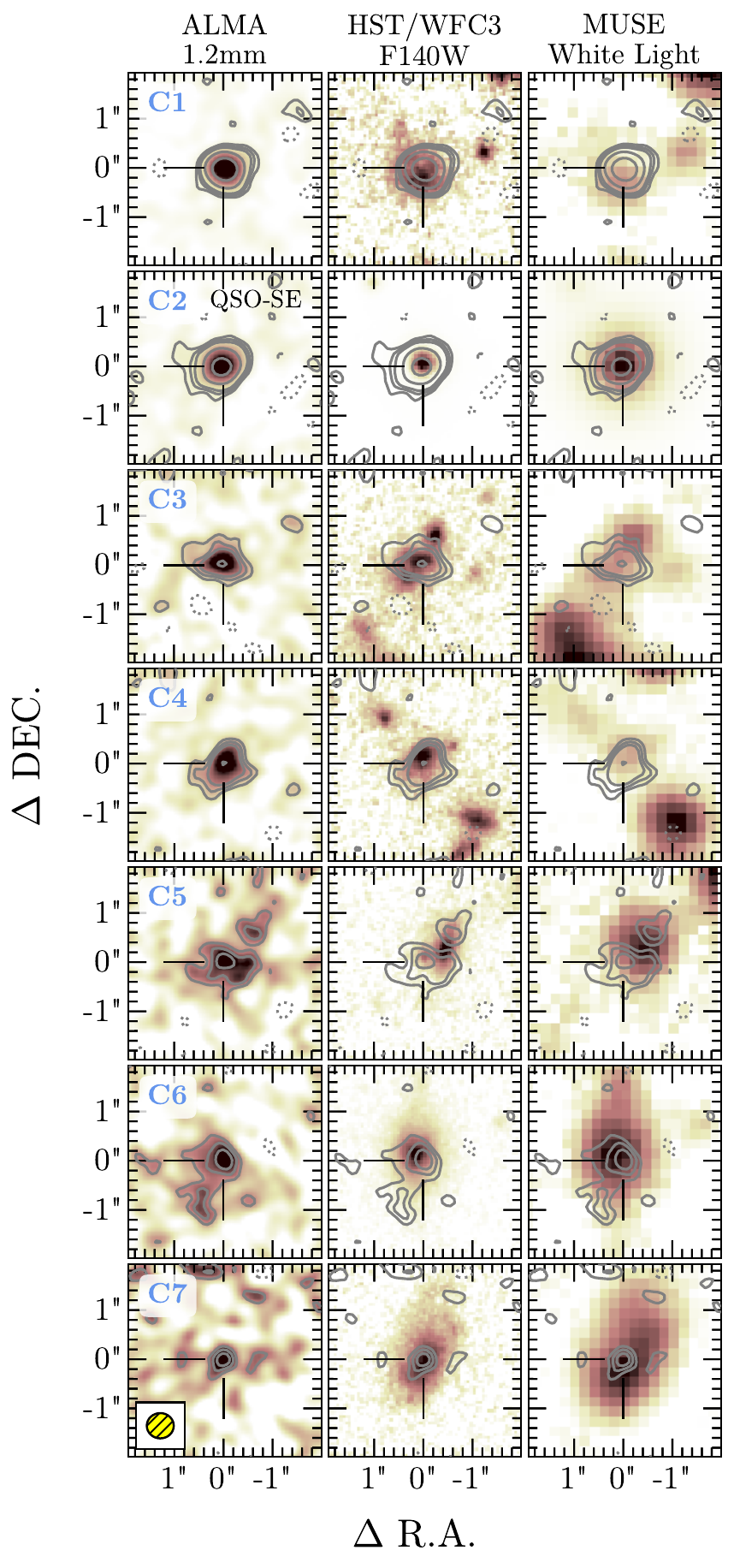}}
       \caption{$5\arcsec\times5\arcsec$ postage stamps of NIR and optical/NUV counterparts of ALMA 1.2-mm continuum-selected sources in the MUDF. {\it From left to right:} 1.2-mm continuum map from ALMA band 6, HST/WFC3 F140W (central observed wavelength $1.4\,{\rm \mu m}$), MUSE white light image (central observed wavelength $655\,{\rm nm}$). The solid contours show the $[2,3, 2^{n}]\times{\rm RMS}$ (with $n\ge2$) levels of the ALMA continuum. The yellow ellipse at the corner of the bottom left panel shows the synthesized ALMA beam.}
       \label{fig:counterparts}
   \end{figure}

In Fig.~\ref{fig:counterparts}, we show postage stamps of the seven ALMA continuum-selected sources and their NIR and optical/NUV counterparts in the field. As also described in Sect.~\ref{ssect:cont_selected_sources}, we cross-matched all the sources detected in the continuum at 1.2 mm with ALMA with the catalog of HST-selected galaxies. Within a projected separation of $0\rlap{.}{\arcsec}5$ in the sky, we found an unambiguous counterpart for all ALMA detections, except for C5. By visually inspecting the HST/WFC3 F140W image of this source, we found that the peak of the ALMA continuum appears to correspond to a faint and compact source located southeast relative to an extended object. Interestingly, the 1.2-mm continuum emission seems to be spatially associated with diffuse emission revealed by HST around the extended galaxy, which suggests irregular morphology, possibly due to the interaction with the nearby compact source. Alternatively, gravitational lensing may represent a viable explanation to the observed emission. However, given the limited $S/N$ of the ALMA data, we cannot confidently assess the nature of this emission. The angular resolution of MUSE observations is insufficient for adequately deblending the spectra of these two closely spaced sources (see, Fig.~\ref{fig:counterparts}). By inspecting the MUSE spectrum extracted at the location of the galaxies, we assigned to both the sources a low-quality redshift estimate of $z\approx 2.4723$ on the basis of only the tentative detection of \ion{C}{IV} $\lambda\lambda1548, 1550$ absorption line feature. Future deep spectroscopic observations using, for example, JWST, are needed to shed light on the characteristics of this system.

A high-quality spectroscopic redshift is constrained from the MUSE spectra for C2, C6, and C7. The remaining ones (i.e., C1, C3, and C4), which are also the faintest in the F140W filter among the counterparts (see, Table~\ref{tbl:1.2mm_cand}), do not have a secure redshift measurement from MUSE or HST grism spectra. The redshifts of the latter, are accurately measured from the CO(3--2) lines detected in the band 3 spectral scans as previously described in Sect.~\ref{ssect:search_spectral_scans}.

   \begin{figure*}[!htbp]
    	\centering
   	\resizebox{\hsize}{!}{
	\includegraphics{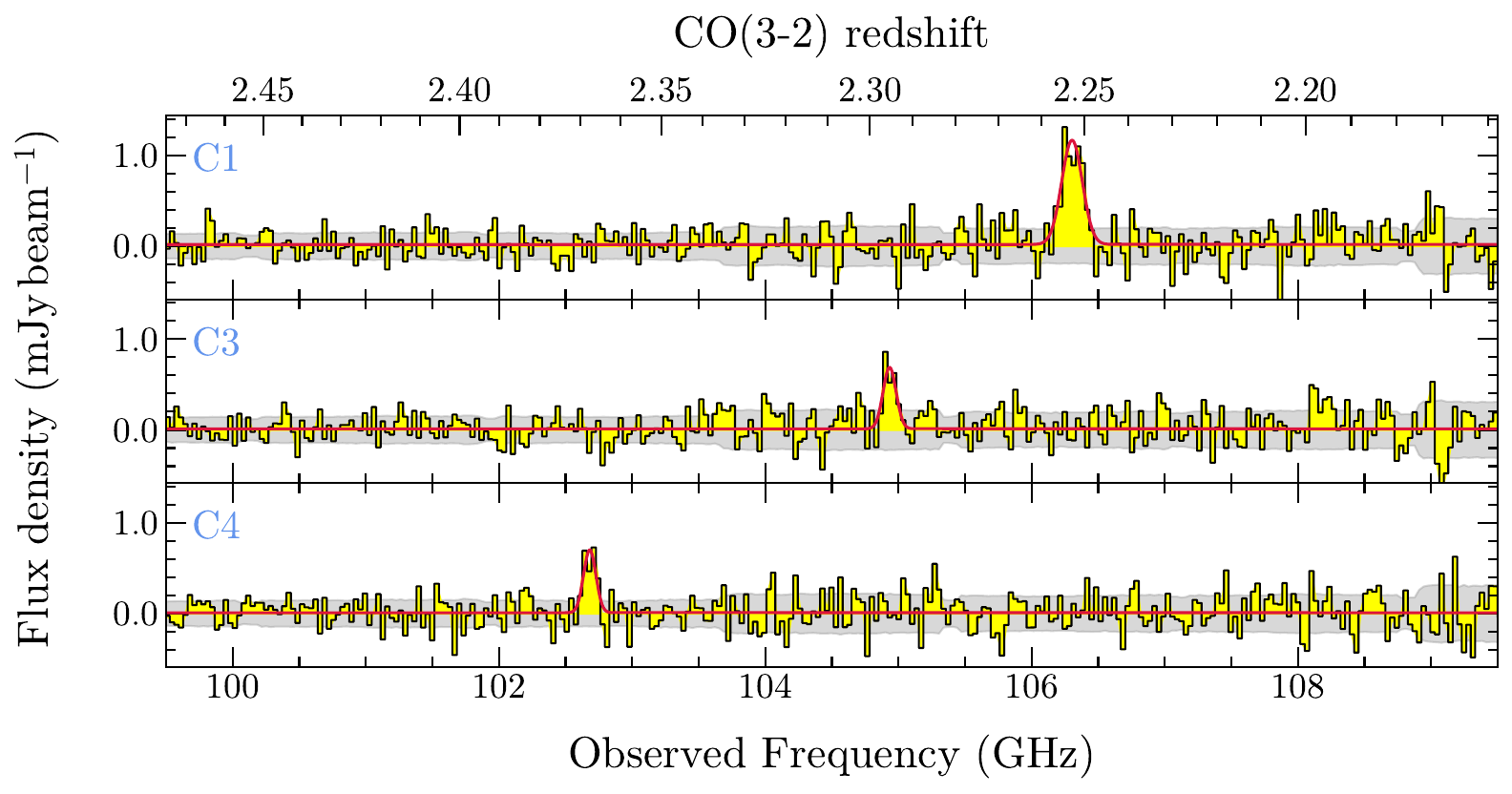}}
       \caption{Spectra extracted from ALMA band 3 spectral scans toward 1.2-mm continuum-selected sources C1, C3 and C4 (from the top to the bottom panel). Data and the best-fit model are shown in yellow and red, respectively. The $\pm1\sigma$ RMS noise range of the data is indicated by the gray shaded area. The top scale reports the expected CO(3--2) line redshift depending on the observed frequency.}
       \label{fig:spectra}
   \end{figure*}
   
\subsection{Source fluxes and luminosities}
\label{ssect:source_fluxes}
All seven ALMA-selected sources detected in the continuum at 1.2 mm appear, at best, marginally resolved. To maximize the recovered signal, we estimated their flux density in the $\sim2{\arcsec}$ tapered image through a single-pixel measurement of the continuum source flux peak that we then rescaled for the PB response at the corresponding location. We then applied the same PB rescaling to the RMS (in ${\rm mJy\,beam^{-1}}$) measured on the 1.2-mm continuum image to estimate the $1\sigma$ uncertainty on $F_{1.2\,{\rm mm}}$. We report these quantities in Table~\ref{tbl:1.2mm_cand}. 

We measured the CO(3--2) line flux of C1, C3, and C4 detected in the band 3 cubes by fitting the spectra extracted from the source peak employing a single Gaussian plus a constant to model the line and the continuum. We then estimated the best-fit model parameters and uncertainties from the posterior probability distributions obtained through the Python Markov chain Monte Carlo (MCMC) ensemble sampler  \texttt{emcee} package \citep{Foreman+2013}, assuming box-like priors based on a visual inspection of the data. We finally derived the line luminosities as \citep[see, e.g.,][]{Solomon+1997}:

\begin{align}
\label{eq:lum}
L_{\rm CO} [L_{\astrosun}] =&  1.04\times 10^{-3}\,S\Delta\varv\,\nu_{\rm obs}\,D_L^2, \\
\label{eq:lum_prime}
L'_{\rm CO} [{\rm K\,km\,s^{-1}\,pc^2}] =& 3.25\times10^7\,S\Delta\varv\,\frac{D_L^2}{(1+z)^3\,\nu_{\rm obs}^2,}
\end{align}

where $S\Delta\varv$ is the velocity-integrated line flux in ${\rm Jy\,km\,s^{-1}}$, $\nu_{\rm obs}$ is the observed central frequency of the line in GHz, $z$ is the source redshift measured from the centroid of the CO line, and $D_L$ is the luminosity distance in Mpc. The relation between Eq.~(\ref{eq:lum}) and (\ref{eq:lum_prime}) is $L_{\rm CO}=3\times10^{-11}\,\nu_{\rm rest}^3\,L_{\rm CO}'$, where $\nu_{\rm rest}$ is the line rest frequency in GHz. In Fig.~\ref{fig:spectra}, we show the spectra and the best-fit models and report the line measurements in Table~\ref{tbl:co32_line}.

\begin{table}[!t]
\def\arraystretch{1.15}
\caption{ALMA CO(3--2) line emitter.}          
\label{tbl:co32_line}      
\centering 
\resizebox{\hsize}{!}{
\begin{tabular}{lccc}  
\toprule\toprule
ID$_{\rm 1.2mm}\,^{(1)}$ & C1 & C3 & C4\\
\cmidrule(lr){1-4}
$z_{\rm CO}\,^{(2)}$ & $2.2529^{+0.0003}_{-0.0003}$ & $2.2954^{+0.0005}_{-0.0005}$ & $2.3676^{+0.0003}_{-0.0003}$  \\
$S\Delta\varv\,^{(3)}\,({\rm Jy\,beam^{-1}\,km\,s^{-1}})$ & $0.63^{+0.06}_{-0.06}$ & $0.24^{+0.06}_{-0.06}$ & $0.23^{+0.04}_{-0.04}$ \\
${\rm FWHM}\,^{(4)}\,({\rm km\,s^{-1}})$ & $510^{+50}_{-50}$ & $330^{+120}_{-80}$ & $320^{+50}_{-40}$ \\
$L_{\rm CO}\,^{(5)}\,(10^{7}\,L_{\astrosun})$ & $2.3^{+0.2}_{-0.2}$ & $0.9^{+0.2}_{-0.2}$ & $0.95^{+0.14}_{-0.14}$  \\
$L'_{\rm CO}\,^{(6)}\,({\rm 10^{10}\,K\,km\,s^{-1}\,pc^2})$ & $1.77^{+0.17}_{-0.17}$ & $0.70^{+0.18}_{-0.18}$ & $0.72^{+0.11}_{-0.10}$\\
\bottomrule
\end{tabular}
}
\tablefoot{Derived properties from the CO(3--2) line.  
 $^{(1)}$Identifier of ALMA 1.2-mm continuum-selected galaxies. $^{(2)}$Redshift estimate for the CO(3--2) line. All reported uncertainties correspond to $1\sigma$ confidence interval as derived from the 16th and 84th percentile of the posterior probability distribution. $^{(3)}$Velocity-integrated line flux. $^{(4)}$Full Width at Half Maximum of the line. $^{(5,6)}$Line luminosities as derived from Eq.~(\ref{eq:lum}) and (\ref{eq:lum_prime}). }
\end{table}

\section{Linking galaxies to absorbers in multiwavelength datasets}
\label{sect:results_discussion}
Absorption studies toward background quasars typically target galaxies in their rest-frame optical/UV. So far, (sub-)millimeter studies (e.g., with ALMA) have focused on specific classes of absorbers, such as high-column-density DLAs \citep[$N_{\ion{H}{I}}\ge10^{20.3}{\rm cm^{-2}}$; see, e.g.,][]{Neeleman+2017, Kanekar+2018, Kaur+2022, Kaur+2022b}. However, little is known about the contribution of dusty and gas-rich galaxies to the cosmic metal enrichment across different epochs of the Universe from untargeted observations \citep{Dudzeviciute+2021}. In this work, we capitalize on our blind ALMA survey to study the association between cosmic gas traced by quasar absorption systems in the MUDF, and galaxies detected at millimeter wavelengths in their dust continuum.

    \begin{figure*}[!t]
    	\centering
   	\resizebox{\hsize}{!}{
	\includegraphics{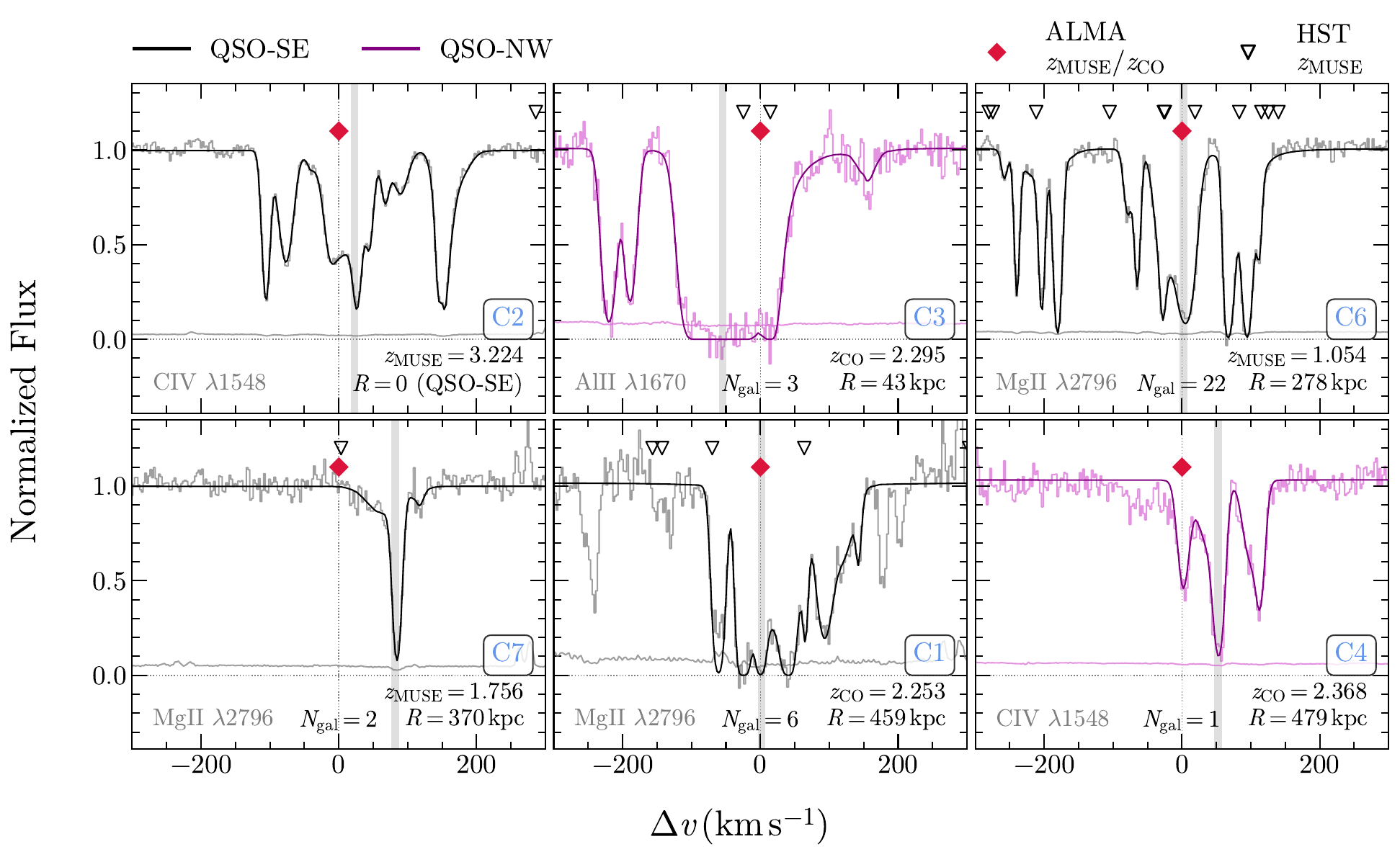}}
       \caption{Connection between ALMA-selected galaxies and absorption systems in the spectra of the QSO-SE (black) and -NW (purple). The line-of-sight separations are centered on the redshift of the ALMA-selected galaxies. The vertical gray band indicates the velocity at which the optical depth of the absorption is $50\%$ of the total ($\tau_{50}$). The transitions shown (\ion{C}{IV} $\lambda1548$, \ion{Al}{II} $\lambda1670$, and \ion{Mg}{II} $\lambda2796$) are limited to the strongest lines among the multiple metals associated to each galaxy. Galaxies selected with ALMA are indicated by red diamonds and ordered (from the top left to the bottom right panel) for increasing sky-projected separation ($R$) from the QSO sightline along which the absorption is detected. HST-selected galaxies linked to the same absorbers are shown with black triangles. The number of  galaxies ($N_{\rm gal}$) associated to each group (see, Sect.~\ref{ssect:groups}) is also reported.}
       \label{fig:abs_line}
   \end{figure*}

\subsection{Gas-galaxy association rate}
With accurate redshift measurements ($\Delta\varv\lesssim 60\,{\rm km\,s^{-1}}$) of our ALMA-selected galaxies from either MUSE spectra or ALMA CO(3--2) line (see Sect.~\ref{ssect:hst_counterparts}, ~\ref{ssect:source_fluxes}), we matched these galaxies with the absorbers in velocity space in either of the quasar sightlines (see, Sect.~\ref{ssect:qso_absorbers}). Following the approach typically adopted in the literature for optical studies \citep[see, e.g.,][]{Dutta+2021, Schroetter+2021, Galbiati+2023, Beckett+2024}, we link the absorbing gas with galaxies by searching for association within $\abs{\Delta\varv}\le 500\,{\rm km\,s^{-1}}$ with respect to the absorbers. We report our results in Fig.~\ref{fig:abs_line}. We found that $100\%$ of our ALMA-selected galaxies -- for which robust spectroscopic redshifts are available -- are unambiguously associated with metal absorbers. This detection rate decreases to $\ge85\%$ when considering the uncertainties related to source C5 (see Sect.~\ref{ssect:hst_counterparts} for details). Indeed, if we adopt the tentative redshift of C5 measured from the low-quality MUSE spectrum ($z\approx 2.47$), we find no connected absorbers for this source. For comparison, we found that only $\approx 25\%$ of all spectroscopically confirmed HST-selected galaxies with $\log M_\star/M_{\astrosun}\ge 8$ are associated with absorbers, increasing to $\approx 45\%$ considering galaxies with $\log M_\star/M_{\astrosun}\ge 10$. Therefore, the association rate for galaxies selected in their rest-frame optical/UV is much lower than that observed for on-average more massive galaxies selected with ALMA in their dust continuum (see, Sect~\ref{ssect:gal_prop}).

Interestingly, we note that three out of the six sources (i.e., C1, C3, C4) for which we find a robust association to absorbers would have been missed if only relying on MUSE and HST identifications. Indeed, such sources are the faintest among the NIR/optical counterparts (see Table~\ref{tbl:1.2mm_cand}) and lie in the MUSE redshift desert. This implies that studies limited to optical wavelengths may miss a significant fraction of the galaxy population around metal absorbers, as also highlighted in previous works \citep[see, e.g.,][]{Neeleman+2017, Neeleman+2018, Neeleman+2019, Fynbo+2018, Galbiati+2024}. Moreover, as the case of C4 exemplifies, ALMA is critical to identify the only galaxy we can associate with the QSO-NW absorber at $z=2.368$, as no other galaxies were previously known at this redshift through MUSE and HST (see, Fig.~\ref{fig:abs_line}). 

We evaluated the number density of our ALMA-selected galaxies by computing the completeness-corrected 1.2-mm number counts and found that these are consistent with that derived in blank field from ASPECS \citep[][]{Gonzalez-Lopez+2020}. In light of the above considerations, we conclude that, since obscured massive galaxies selected with ALMA have sufficient number density to be relevant in absorption line studies, blind (sub-)millimeter follow-ups of optical/NIR surveys are thus essential in providing the most complete view of the galaxy population residing around absorbers.

\subsection{Properties of the ALMA galaxies around absorbers}
\label{ssect:gal_prop}
In Fig.~\ref{fig:main_seq}, we compare the stellar mass and SFR of the dusty galaxies detected with ALMA (see Appendix~\ref{app:sed} for full details) to the rest of the optical/UV-selected galaxy population in the MUDF and the sample of submillimeter galaxies (SMG) from \citet{Dudzeviciute+2020} at $z\sim 1-2.5$. The ALMA sources lie at the high-mass end of the stellar mass function at those redshifts and are among the most massive members of the groups in which they reside (see Sect.~\ref{ssect:groups}). On the other hand, our ALMA galaxies appear to be less massive, but within in the scatter of the population of coeval SMGs, lying on the asymptotic part of the main sequence at $z\simeq 1-2$ \citep[see,][]{Popesso+2023}, although close to the starburst region.

       \begin{figure}[!t]
    	\centering
   	\resizebox{\hsize}{!}{
	\includegraphics{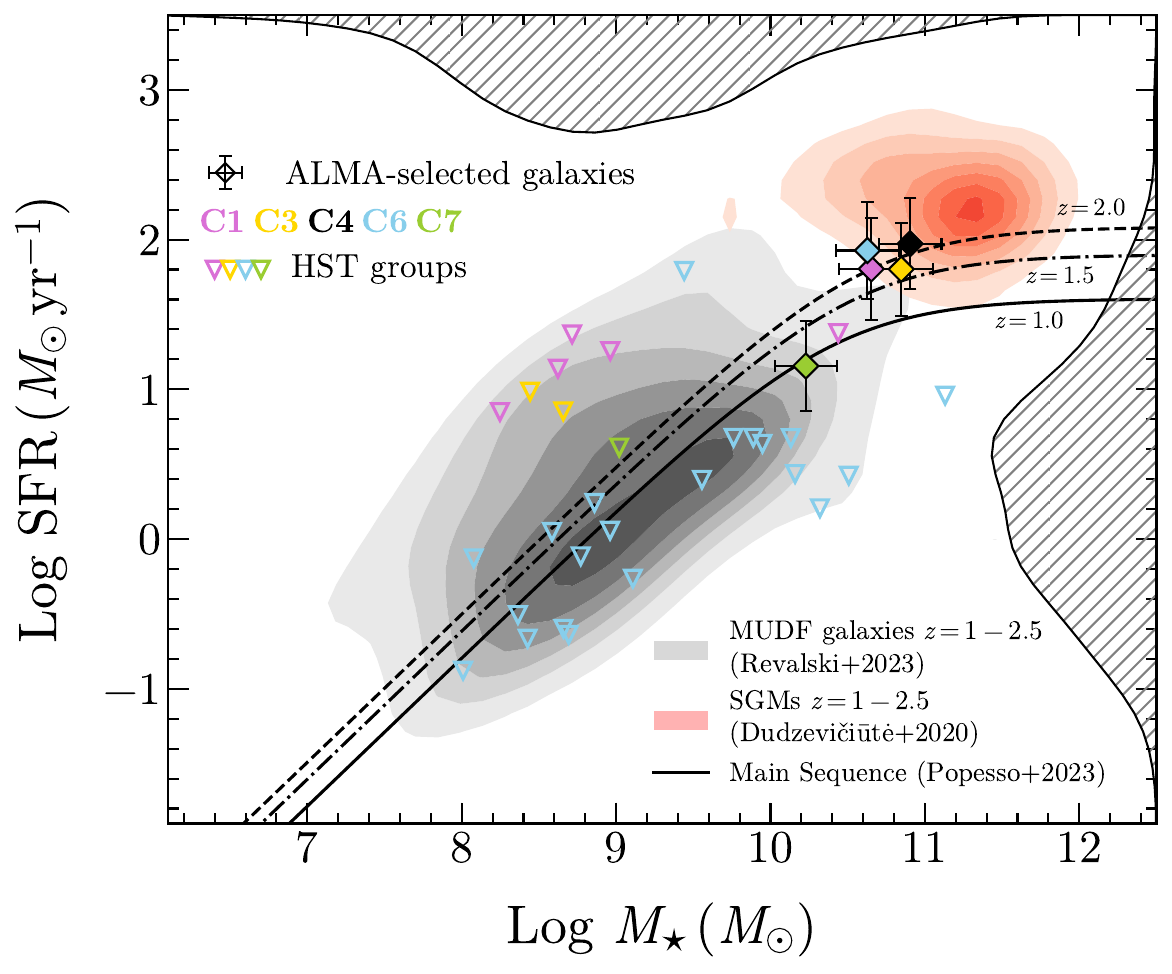}}
       \caption{Star formation rate-stellar mass diagram of galaxies detected in the MUDF. The gray density plot shows the distribution of all galaxies selected with HST in the rest-frame optical/UV at $z\sim 1-2$. The marginalized distributions of $M_\star$ and SFR are also shown on the top and right axis, respectively. The red density plot report the sample of SMG with photometric redshift within $1\le z \le 2.5$ from \citet{Dudzeviciute+2020}. The ALMA-selected galaxies with properties derived from SED modeling are reported with diamonds. Colored triangles mark the HST-selected group galaxies associated with the same absorption system of each ALMA-selected galaxy. The star-forming main sequence at $z\sim 1-2$ from \citet{Popesso+2023} is also shown.}
       \label{fig:main_seq}
   \end{figure}
   
Using our measurements of the galaxy flux density at 1.2 mm on the Rayleigh-Jeans tail of the dust SED, we provide rough estimates of the dust masses ($M_{\rm dust}$) of ALMA continuum-selected galaxies (excluding C5 for which we do not have reliable redshift measurement, and C2 that is the QSO-SE). By following a typical approach adopted in (sub-)millimeter studies of high-$z$ galaxies \citep[e.g.][]{Casey+2012, Riechers+2013, Faisst+2020}, we employ a single modified black-body emission assuming a characteristic dust temperature of $T_{\rm dust}=32\,{\rm K}$, and emissivity index $\beta=1.8$ typical of SMGs \citep[see][]{Dudzeviciute+2020}, and we find values in the range $\log\,(M_{\rm dust}/M_{\astrosun})\simeq 8.0-8.6$. Using the same assumptions, we find a lower value of $\log\,(M_{\rm dust}/M_{\astrosun})\simeq 7.8$ for C7, which appears to be more similar to typical main-sequence galaxies. By integrating the black-body models we computed the IR luminosities and converted them to ${\rm SFR}_{\rm IR}$ using the calibration from \citet{Murphy+2011}, and find values in the range $\log\,({\rm SFR}_{\rm IR}/M_{\astrosun}\,{\rm yr^{-1}})\simeq1.6-2.4$, comparable to those we derived from galaxy SED modeling (see Table~\ref{tbl:sed_fit}). 

From the CO(3--2) emission line in C1, C3, C4, and the \ci{609} line detected in C6 (see, Sect.~\ref{ssect:line_selected_sources}), we derive estimates of galaxy molecular gas fractions. For C1, C3, and C4 we adopt a CO(3--2)-to-CO(1--0) conversion factor of $r_{31}=0.80\pm0.14$ from \citet{Boogaard+2020} and a range of $\alpha_{\rm CO}$. At the lower end, we use a value of $0.8\,M_{\astrosun}/({\rm K\,km\,s^{-1}\,pc^2)}$, typical of SGMs and quasar hosts \citep[ see, e.g.,][]{Downes+1998}. At the upper end, we adopt  $4.3\,M_{\astrosun}/({\rm K\,km\,s^{-1}\,pc^2)}$ as measured in giant molecular clouds in the Milky Way \citep[see, e.g.,][]{Bolatto+2013, CarilliWalter2013}. For C6, we first derive atomic carbon mass from equation 2 in \citet{Weiss+2005} and we convert it to molecular gas mass by assuming a C-to-H$_2$ abundance ratio of $X{\rm [CI]}/X{\rm [H_2]}=M_{\rm CI}/(6M_{\rm H_2})=(8.4\pm3.5)\times10^{-5}$, as derived by \cite{Walter+2011} for a sample of $z\sim 2-3$ FIR-bright sources. By employing such recipes, we find $f_{\rm gas}=M_{\rm H_2}/(M_{\rm H_2}+M_\star)=51^{+14}_{-17}\%$, $24^{+15}_{-10}\%$, $20^{+12}_{-9}\%$, and $65^{+10}_{-12}\%$ for C1, C3, C4, and C6 respectively. These values are scattered in the typical range observed in both main-sequence galaxies and ULIRGs/SMGs at $z\gtrsim1$ \citep[see, e.g.,][]{Tacconi+2018, Birkin+2021}. Interestingly, we note that C1 and C6 
exhibit high values of the gas fraction and also live in rich groups (see, Sect.~\ref{ssect:groups}). This fact might suggest that efficient gas accretion is in place in these structures.

\subsection{The link between the large-scale galaxy environment and the absorbing gas}
\label{ssect:groups}
\begin{figure}[!t]
    	\centering
   	\resizebox{\hsize}{!}{
	\includegraphics{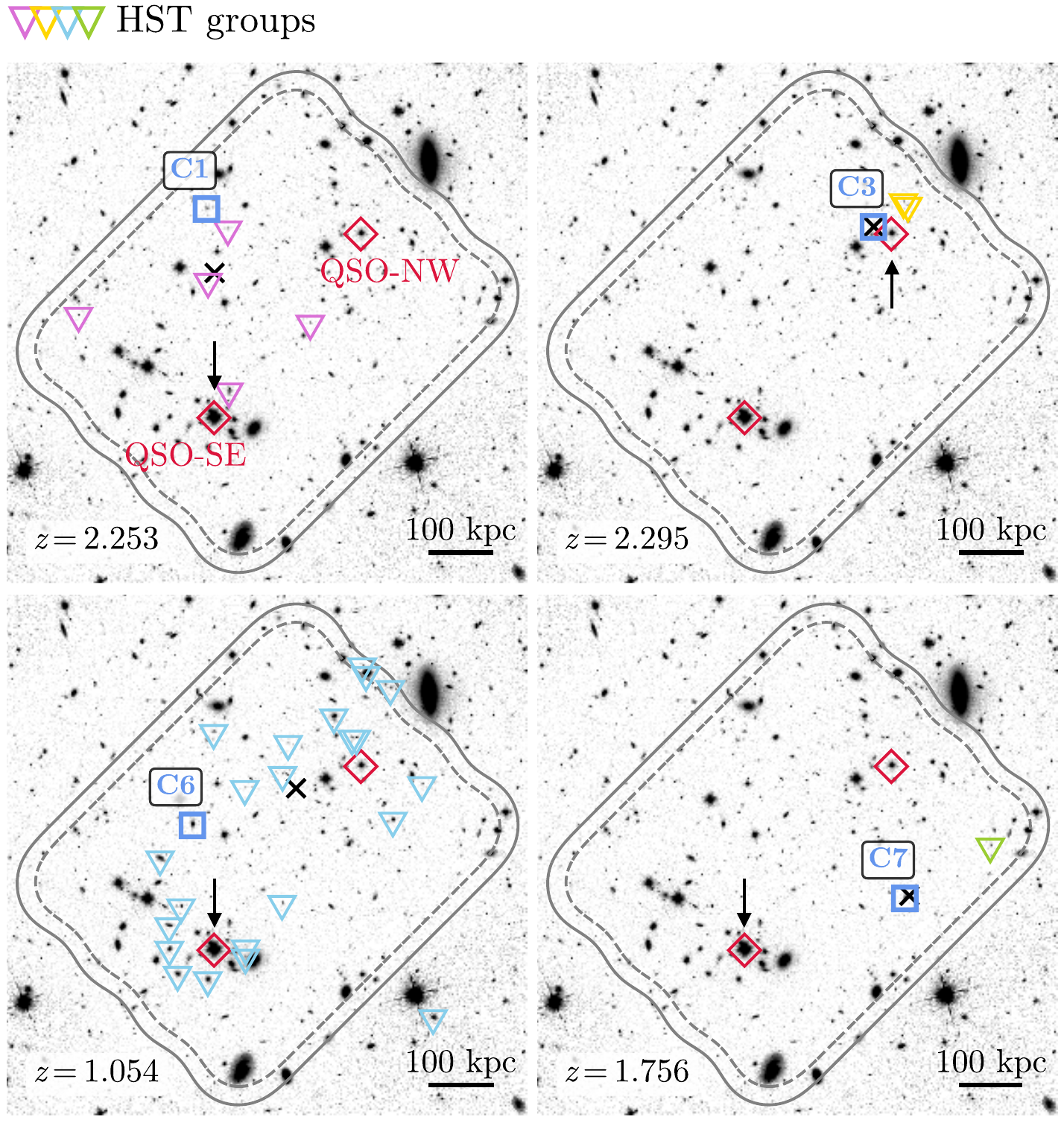}}
       \caption{Groups of galaxies around ALMA sources (blue squares). The colored triangles indicate the HST-selected galaxies belonging to the same group found around the redshift of ALMA sources (bottom left corner). The two quasars in the MUDF are labeled with red diamonds, the black arrow indicates the QSO sightline in which absorption at the redshift of each group is detected. The black crosses mark the projected location of the galaxy stellar mass-weighted barycenter of the groups. The gray solid and dashed lines report the PB response at 0.3 and 0.5, respectively, of the ALMA band 6 mosaic.}
       \label{fig:hst_groups}
   \end{figure}

In Sect.~\ref{ssect:hst_cat}, we found that many HST-selected sources do not live in isolation but are associated with galaxy groups instead. In Fig.~\ref{fig:hst_groups}, we show that four (out of six) of our dusty galaxies (i.e., C1, C3, C6, and C7) detected with ALMA actually live in such overdense regions\footnote{From the total stellar mass of the groups we estimated the dark matter halo masses of the systems in which ALMA galaxies reside by employing the stellar-to-halo mass relation from \citet{Moster+2010}, and obtain an average halo mass of $\sim 10^{13}\,M_{\astrosun}$. This is comparable to previous estimates of the halo mass of SMGs at $z>1$ from clustering analysis \citep{Hickox+2012, Wilkinson+2017, Lim+2020, Stach+2021}.}. This is also evident from Fig.~\ref{fig:abs_line}, in which multiple galaxies selected in HST are found to be associated with the same absorbers as those identified around ALMA galaxies. The high association rate between galaxies in groups and absorbers has also been found in previous works \citep[see, e.g.,][]{Nielsen+2018, Dutta+2020, Hamanowicz+2020, Lofthouse+2023, Qu+2023}, suggesting that the large-scale environment plays a significant role in shaping the absorption properties. Indeed, these absorbers exhibit multiple components and are observed to be strong, showing typical rest-frame equivalent widths of ${\rm EW}_{\rm CIV} \gtrsim 0.3\,\AA{}$, ${\rm EW}_{\rm MgII}\gtrsim 1\,\AA{}$, and with complex kinematics. In particular, we note that dust-mass-selected sources living in groups of more than two galaxies (i.e., C1, C3, and C6) are associated to absorbers exhibiting high column density of $N_{\rm AlII,\,MgII}\apprge 10^{16}\,{\rm cm^{-2}}$, while C4, and C7 show significantly lower values of $N_{\rm CIV,\,MgII}\simeq 10^{13}-10^{14}\,{\rm cm^{-2}}$. Other works which studied quasar absorption around galaxy groups at different redshifts reported similar evidence \citep[see, e.g.,][]{Lan+2014, Fossati+2019, Dutta+2021, Nelson+2021, Galbiati+2023, Galbiati+2024}. 

We now turn to investigate the 3D separation between galaxies and the associated gas structures. The dusty galaxies selected with ALMA are more aligned in redshift space with the center of absorption features than the other connected galaxies. In particular, we find that C1 and C6, which are embedded in rich groups, are located in correspondence of the $50\%$ of the optical depth ($\tau_{50}$) of the associated absorbers (see, Fig.~\ref{fig:abs_line}). Similarly, despite living in less overdense regions, we find that the remaining ALMA galaxies (i.e., C3, C4, and C7) lie within $100\,{\rm km\,s^{-1}}$ from $\tau_{50}$. This fact, in combination with their massive nature (see, Sect.~\ref{ssect:gal_prop}), suggests that these galaxies sit close to the center of the potential well of their large-scale structures. While our ALMA-selected galaxies show preferential alignment with the quasar absorbers along the line of sight, this is not necessarily the case if we consider the sky-projected distances from the location of the quasars. The impact parameters are typically large, of the order of $\sim 200-400\,{\rm kpc}$ (see, Fig.~\ref{fig:hst_groups}), compared to the other galaxies associated to the absorbers. Therefore, while (sub-)millimeter observations are critical to gain a complete view of the galaxy population and possibly the group potential well, our findings suggest that the absorbing gas may not be located in the inner CGM of the dusty galaxies but is still retained within a few virial radii\footnote{The virial radius for a halo mass of $10^{13}\,M_{\astrosun}$, typical of the ALMA-selected galaxies \citep{Stach+2021} according to \citet{Moster+2010}, is $\approx310$ and $220\,{\rm kpc}$ at $z=1$ and 2, respectively.}.
In this regard, massive dusty galaxies appear to be more tracers of the large scale structure than the actual hosts of the absorbers. However, we caution that this may also be a consequence of the low number density of these galaxies. More SMG-QSO pairs at small separations are needed to statistically assess our conclusions. We note that only one out of six detections, that is the source C3, lies both close to the center of the associated absorber in redshift space and at low impact parameter ($R=43\,{\rm kpc}$) from QSO-NW sightline, which points to a significant contribution from the CGM of this galaxy to the observed absorption profile. However, given that C3 belongs to a compact system of three sources all located within $100\,{\rm kpc}$, we cannot exclude that hydrodynamical and gravitational interactions may play an important role in shaping the properties of the surrounding gas \citep[][]{Gunn+1972, Merritt+1983, Boselli+2006, Klitsch+2019}. 

\begin{figure}[!t]
    	\centering
   	\resizebox{\hsize}{!}{
	\includegraphics{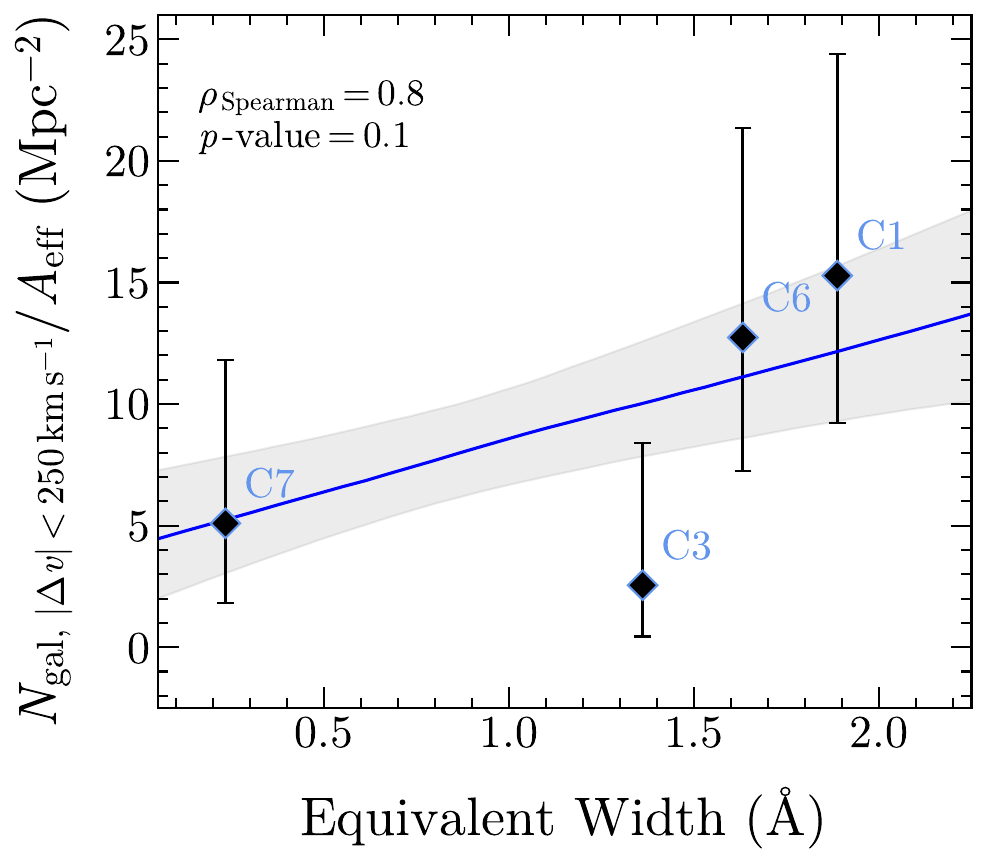}}
       \caption{Galaxy surface density in groups as a function of EW of the associated absorbers (\ion{Mg}{II} for C1, C6, and C7, and \ion{Al}{II} for C3, respectively). Number density is computed by counting the number of galaxies ($N_{\rm gal}$), and dividing for the effective area ($A_{\rm eff}$) as described in Sect.~\ref{ssect:groups}. The associated uncertainties are obtained from Poisson statistics for low number counts \citep{Gehrels1986}. The blue solid line and gray shaded area are the best fit linear model ($N_{\rm gal}/A_{\rm eff}= a\times{\rm (EW/\AA)}+b\,{\rm Mpc^{-2}}$, with $a=4\pm2$, $b=4\pm3$) and $1\sigma$ confidence interval, respectively. The Spearman's rank correlation coefficient and the corresponding $p$-value are also reported.}
       \label{fig:sigma_ew}
   \end{figure}

In light of the aforementioned considerations, the observed absorption profiles associated to groups could potentially arise from the superposition of the halos of multiple galaxies located around the massive sources identified with ALMA \citep[see, e.g.,][]{Bordoloi+2011, Nielsen+2018, Dutta+2020}. Therefore, one might expect the strength of the absorption to correlate with the projected galaxy surface density \citep[see, e.g.,][]{Galbiati+2024}. To test this hypothesis, we centered on the sky position and redshift of ALMA galaxies C1, C3, C6 and C7, and count the number of HST-selected sources within a velocity window of $\abs{\Delta\varv}\le 250\,{\rm km\,s^{-1}}$ (chosen based on the broadest absorption system, see Fig.~\ref{fig:abs_line}), and encompassed radius corresponding to the impact parameter from the QSO. We then divide the number of galaxies by the projected area corrected for the coverage of the MUSE FoV. In Fig.~\ref{fig:sigma_ew}, we report such galaxy surface density against the EW of the associated absorbers\footnote{We used \ion{Mg}{II} for C1, C6, and C7, and \ion{Al}{II} for C3, since we have no \ion{Mg}{II} detection for this source. These are chosen as they have similar $f_{\rm osc}\lambda_{\rm rest}^{2}$, where $f_{\rm osc}$ is the oscillator strength, and $\lambda_{\rm rest}^{2}$ is the rest-frame wavelength of the transitions. However, we caution that due to the differences in the relative abundances, at fixed EW different ions do not trace the same gas density.}. As a result, we find a hint of positive correlation\footnote{The Spearman's rank correlation test \citep{Spearman1904} yields a degree of monotonic correlation of $\rho=0.8$ with a $p$-value $0.1$, indicating that a positive correlation is in place at $80\%$ confidence level.}, which is consistent with the scenario proposed above. In order to show the potential correlation, we fit the data with a linear model by employing the Python MCMC ensemble sampler \texttt{emcee} package \citep{Foreman+2013} adopting a Poisson maximum likelihood estimator and flat priors on the free parameters. In Fig.~\ref{fig:sigma_ew}, we report the best fit and $1\sigma$ confidence interval. However, we note that our analysis relies on a handful of sources and therefore is limited by the low statistics. Hence, surveys tailored to probe a broader range of galaxy overdensities around absorbers are required to draw more firm conclusions. 

\section{Summary \& Conclusions}
\label{sect:summary_conclusions}
In this work, we presented an ALMA blind dust continuum and molecular gas survey around the sightlines of the $z\approx3.22$ quasar pair in the MUDF. By combining our (sub-)millimeter observations with available multiwavelength data from NIR to UV, we investigated the role of the dusty galaxy population in association to the intervening gas structures revealed in absorption in the quasar spectra. Below, we summarize our main findings:
\begin{itemize}
\item We identified a robust sample of seven sources detected at 1.2-mm in their dust continuum with ALMA. We determined their spectroscopic redshift from either the inspection of their MUSE spectra or via the CO(3--2) line detected with ALMA band 3 spectral scans.
\item We found that the dust-selected galaxies and metal absorbers are associated within $500\rm\,km\,s^{-1}$. This implies that at least $\ge 85\%$ of the dusty galaxies uncovered with ALMA are unambiguously linked to multiple absorption features arising from metal ions and observed in either of the quasar sigthlines. This association rate is significantly higher than in previous surveys at NIR--UV wavelengths.
\item By combining multiband photometry and information about dust and molecular gas gathered from our (sub-)millimeter data, we investigated the nature of the ALMA-selected galaxies by performing SED modeling. As a result, most of these galaxies appear massive ($M_{\star}\simeq 10^{10.2-10.7}\,M_{\astrosun}$), highly star forming (${\rm SFR}\simeq 10^{1.2-2.0}\,M_{\astrosun}\,{\rm yr^{-1}}$), and exhibit high dust and molecular gas content ($M_{\rm dust}\simeq 10^{7.8-8.6}\,M_{\astrosun}$, $f_{\rm H2}\simeq 20 - 65\%$), resembling typical SMGs at similar redshifts.
\item We discovered that a large fraction of ALMA-selected sources (4/6) live in galaxy groups in which they are among the most massive members. While the associated absorbers show complex kinematics, our galaxies are better aligned with the center of the absorption features than the other galaxies previously identified in the optical. This may suggest that these massive and dust-obscured galaxies sit close to the center of the gravitational potential well of their large-scale structures. However, the impact parameters of the ALMA galaxies are typically large, $\sim 200-400\,{\rm kpc}$, indicating that absorption probed by the QSO sightlines might originate within a few virial radii.

\end{itemize}

This study highlights the crucial role of (sub-)millimeter surveys to obtain a complete census of the galaxy population around metal enriched cosmic gas structures revealing the presence of massive and dust-obscured galaxies. However, further efforts are needed to understand which processes are responsible for distributing metals into the IGM and how different galaxy populations contribute to the metal enrichment across the cosmic time. Furthermore, denser surveys are required to statistically investigate in absorption the properties of the inner CGM gas at the interface with the ISM of dusty galaxies. 

\begin{acknowledgements} This paper uses the following ALMA data: ADS/JAO.ALMA\#2021.1.00285.S, ADS/JAO.ALMA\#2023.1.00461.S. ALMA is a partnership of ESO (representing its member states), NSF (USA) and NINS (Japan), together with NRC (Canada), MOST and ASIAA (Taiwan), and KASI (Republic of Korea), in cooperation with the Republic of Chile. The Joint ALMA Observatory is operated by ESO, AUI/NRAO and NAOJ. Based on observations with the NASA/ESA Hubble Space Telescope obtained from the MAST Data Archive at the Space Telescope Science Institute, which is operated by the Association of Universities for Research in Astronomy, Incorporated, under NASA contract NAS5-26555. Support for program Nos. 15637 and 15968 was provided through a grant from the STScI under NASA contract NAS5-26555. These observations are associated with program Nos. 6631, 15637, and 15968. The MUSE portion of this project has received funding from the European Research Council (ERC) under the European Union Horizon 2020 research and innovation program (grant agreement No. 757535) and by Fondazione Cariplo (grant No.2018-2329). This project was supported by the ERC Consolidator Grant 864361 (CosmicWeb) and by Fondazione Cariplo grant no. 2018-2329, 2020-0902. IRS acknowledges support from STFC (ST/X001075/1). This research made use of Astropy\footnote{\url{http://www.astropy.org}}, a community-developed core Python package for Astronomy \citep{AstropyI, AstropyII,AstropyIII}, NumPy \citep{Numpy}, SciPy \citep{Scipy}, Matplotlib \citep{Matplotlib}.
\end{acknowledgements}


\bibliographystyle{aa} 
\bibliography{MyBib_over} 

\begin{appendix}
\section{Spectral Energy Distribution of the ALMA-selected galaxies}
\label{app:sed}
We derived properties of our ALMA-selected galaxies with secure photometric measurements (i.e., excluding C2 that is QSO-SE, and C5; see Sect.~\ref{ssect:hst_counterparts}), such as $M_{\star}$, SFR, and visual extinction ($A_{\rm V}$) by performing simultaneous fit of multiwavelength photometry and MUSE spectra. The photometry includes custom MUSE top-hat filters, covering $4750\,\AA$ to $9300\,\AA$ (as described in table 2 in \citealt{Fossati+2019}), multiple UV to NIR HST filters (WFC3/F140W, F125W, and WFPC2/F450W, F702W, see \citealt{Revalski+2023}), VLT/HAWK-I Ks filter \citep{Revalski+2024}, and 1.2-mm flux from ALMA band 6 observations. 

\begin{figure}[!t]
    	\centering
   	\resizebox{\hsize}{!}{
	\includegraphics{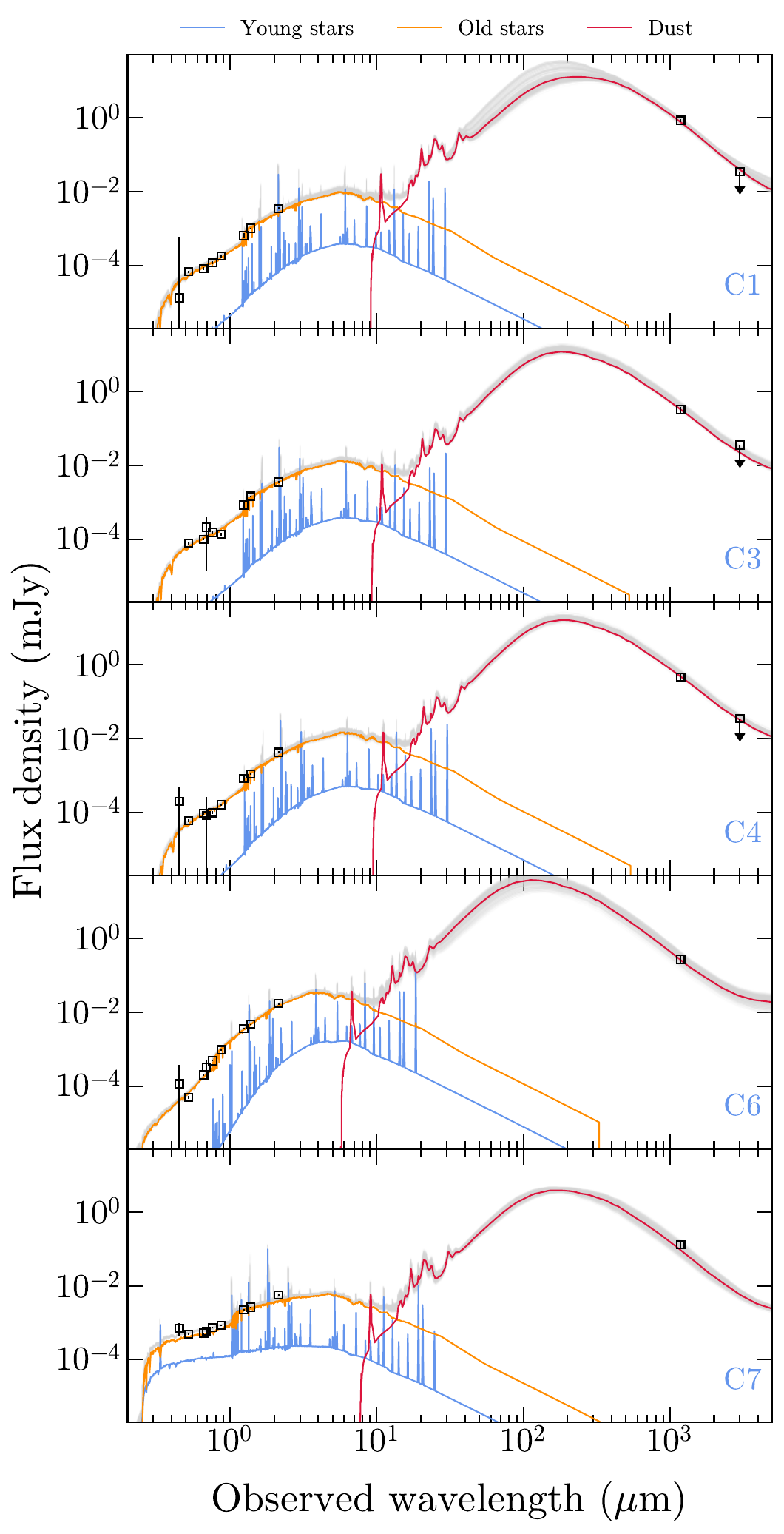}}
       \caption{Best fit SED models of ALMA-selected galaxies. The black squares report the multiband photometric data. Flux upper limits at 5-$\sigma$ for C1, C3, and C4, obtained from 3-mm continuum images are also reported. Red, blue, and orange lines represent the dust emission, and the young (star age $ <10\,{\rm Myr}$) and old ($>10\,{\rm Myr}$) stellar components, respectively. The gray lines show random extraction from the posterior probability distribution of the total galaxy SED.}
       \label{fig:gal_seds}
   \end{figure}

We used the Monte Carlo Spectro-Photometric Fitter ({\sc MC-SPF}, \citealp{Fossati+2018,Fossati+2019b}) code. We built a grid of stellar spectra based on the updated version of the \citet{Bruzual+2003} high-resolution models at $40\%$ solar metallicity, assuming a \citet{Chabrier2003} initial mass function and a delayed exponentially declining star formation history\footnote{The set of models is publicly available at \url{http://www.bruzual.org/CB19/}. A detailed description is provided in appendix A of \citet{Sanchez+2022}.}. We scaled a grid of nebular emission lines models from \citet{Byler+2018} to the number of Lyman continuum photons from the stellar models, converted it into flux at the redshift of the galaxies and then added to templates. To include the effect of dust attenuation on both the stellar continuum and the nebular emission lines, we assumed the double attenuation law from \citet{Calzetti+2000}, which accounts for extra extinction of the young stars compared to those older than $10\,{\rm Myr}$. During the fitting procedure, we impose an energy balance between the UV flux from young stars that is absorbed by dust grains and re-emitted at FIR wavelengths. We used the dust emission model from \citet{Dale+2014}. We report best-fit results from SED modeling in Fig.~\ref{fig:gal_seds} and Table~\ref{tbl:sed_fit}. 

\begin{table}[!t]
\def\arraystretch{1.15}
\caption{Results from SED fitting of the ALMA-selected galaxies.}         
\label{tbl:sed_fit}      
\centering 
\resizebox{0.9\hsize}{!}{
\begin{tabular}{lccc}
\toprule\toprule
ID$_{\rm 1.2mm}\,^{(1)}$ & ${\rm log} \,M_{\star}\,^{(2)}$ & ${\rm \log SFR}\,^{(3)}$ & $A_V\,^{(4)}$\\
	 	& ($M_{\astrosun}$) & ($M_{\astrosun}\,{\rm yr^{-1}}$) &(mag) \\
\cmidrule(lr){1-4}
C1& $10.7\pm0.2$& $1.8\pm0.3$& $1.86\pm0.18$\\
C3 & $10.8\pm0.2$& $1.8\pm0.3$& $1.77\pm0.08$\\
C4& $10.9\pm0.2$& $2.0\pm0.3$& $1.99\pm0.06$\\
C6& $10.6\pm0.2$& $1.9\pm0.3$& $3.25\pm0.14$\\
C7& $10.2\pm0.2$& $1.2\pm0.3$& $0.58\pm0.02$\\

\bottomrule
\end{tabular}
}
\tablefoot{$^{(1)}$Identifier of ALMA 1.2-mm continuum-selected galaxies. $^{(2)}$Logarithm of the stellar mass. $^{(3)}$Logarithm of the star-formation rate. $^{(4)}$Visual extinction. Statistical uncertainties on measurements are estimated from the 16th and 84th percentiles of the parameter posterior probability distribution. Uncertainties on $M_\star$ and SFR are reported after adding in quadrature 0.3 and 0.2 dex, respectively, to account for systematics \citep[see,][]{Pacifici+2023, Beckett+2024}.}
\end{table}

\end{appendix}

\end{document}